\documentclass[useAMS,usenatbib]{mn2e}

\usepackage{comment}
\usepackage{amsmath}
\usepackage{amssymb}
\usepackage{graphicx}
\usepackage{booktabs}
\allowdisplaybreaks[1]
\title[The Hall effect in star formation]{The Hall effect in star formation}
\author[C. Braiding and M. Wardle]{C. R. Braiding\thanks{E-mail:
catherine.braiding@gmail.com} and M. Wardle\\
Research Centre in Astronomy, Astrophysics and Astrophotonics\\
Department of Physics and Astronomy, Macquarie University, Sydney, NSW 2109, Australia}
\begin{document}

\date{Modified: 2012 January 18}

\pagerange{\pageref{firstpage}--\pageref{lastpage}} \pubyear{2012}

\maketitle

\label{firstpage}

\begin{abstract}
Magnetic fields play an important role in star formation by regulating the
removal of angular momentum from collapsing molecular cloud cores. Hall
diffusion is known to be important to the magnetic field behaviour at many of
the intermediate densities and field strengths encountered during the
gravitational collapse of molecular cloud cores into protostars, and yet its
role in the star formation process is not well-studied. We present a
semianalytic self-similar model of the collapse of rotating isothermal
molecular cloud cores with both Hall and ambipolar diffusion, and similarity
solutions that demonstrate the profound influence of the Hall effect on the
dynamics of collapse. 

The solutions show that the size and sign of the Hall parameter can change the
size of the protostellar disc by up to an order of magnitude and the
protostellar accretion rate by fifty per cent when the ratio of the Hall to
ambipolar diffusivities is varied between $-0.5 \le \eta_H / \eta_A \le 0.2$.
These changes depend upon the orientation of the magnetic field with respect
to the axis of rotation and create a preferred handedness to the solutions
that could be observed in protostellar cores using next-generation instruments
such as ALMA. 

Hall diffusion also determines the strength and position of the shocks that
bound the pseudo and rotationally-supported discs, and can introduce subshocks
that further slow accretion onto the protostar. In cores that are not
initially rotating (not examined here), Hall diffusion can even induce
rotation, which could give rise to disc formation and resolve the magnetic
braking catastrophe. The Hall effect clearly influences the dynamics of 
gravitational collapse and its role in controlling the magnetic braking and 
radial diffusion of the field merits further exploration in numerical
simulations of star formation. 
\end{abstract}


\section{Introduction}

Low-mass stars form by the gravitational collapse of molecular cloud cores
over many orders of magnitude in size and density. Cores form within molecular
clouds as a result of turbulent fluctuations, and they gradually contract as
ambipolar diffusion erodes the magnetic support \citep[e.g.][]{tt2009,
fetal2010}. Turbulence may also support the core  against collapse and its
decay can aid in triggering star formation \citep[e.g.][]{mk2004, bpetal2007}.
The core becomes unstable when the mass-to-flux ratio exceeds the critical
value 
\begin{equation} 
    \left(\frac{M}{\Phi}\right)_{crit} = \frac{C_\Phi}{G^{1/2}}
\label{mfratio}
\end{equation}
where $C_\Phi$ is a dimensionless (in cgs units) numerical coefficient that
depends upon the magnetic field and density distributions \citep{ms1956,
ms1976, mzgh1993}. 

The core collapses dynamically into what is termed a ``pseudodisc''
\citep{gs1993a, gs1993b}, which has a flattened shape due to the material
falling in preferentially along the magnetic field lines. The field is
effectively frozen into the cloud, and the infalling material is deflected by
the field lines towards the equatorial plane. The pseudodisc contracts
dynamically in the radial direction, dragging the field lines into a split
monopole (hourglass) configuration \citep{gs1993a, gs1993b} that is consistent
with observations of the polarisation of dust continuum emission in cores
\citep{cc2006, grm2006, ggg2008, aetal2009}. The build up in magnetic pressure
acts as an impediment to further collapse, however, the magnetic tension in
the envelope never suffices to suspend the envelope against the gravity of the
growing protostar \citep{asl2003a, asl2003b}. 

Within the pseudodisc ambipolar diffusion becomes important \citep{dm2001},
leading to a decoupling of the magnetic field from the neutral gas, which
takes place within an outwardly-propagating MHD shock \citep{lm1996}. This
shock (referred to as the ``magnetic diffusion shock'' in this work) is a
continuous transition in the magnetic field and density \citep{kk2002},
inwards of which the neutral material falls in at a near-free fall speed due
to the reduced magnetic support. The centrifugal force becomes important and
triggers the formation of a hydrodynamic shock that strongly decelerates the
infalling matter and allows a Keplerian disc to form \citep{sal1987}. A disc
wind or jet may form, launched from the inner regions of the collapse
\citep{k1989, t2002, asl2003b}. 

Angular momentum is removed from the pseudodisc by the twisting of magnetic
field lines, which transport angular momentum from the inner parts of the core
towards its outer regions \citep{bm1994}. The amount of magnetic braking
affecting the collapse, and hence the existence and sign of the Keplerian
protostellar disc, is determined by the coupling of the field to the charged
particles and the drift of these against the neutral component in response to
the electric field in the neutral rest frame. The Lorentz force is transmitted
to the neutral gas through the drag forces caused by collisions between the
neutral and charged particles \citep[e.g.][]{ks2011}. 

Simulations of star formation typically approximate the magnetic field
behaviour by ideal magnetohydrodynamics (IMHD), where the mass-to-flux ratio
is held constant and the magnetic field is regarded as being frozen into the
neutral medium \citep[e.g.][]{glsa2006, ml2008, mmi2008}. In this situation
the magnetic field and the particles move together in the collapsing flow,
however this simplification only truly applies in the outermost regions of
gravitational collapse where the density is low. If IMHD were to hold true
throughout the collapse the magnetic flux in the star would be $10^3$--$10^5$
times larger than that observed in young stars \citep[this is the ``magnetic
flux problem'', described in][]{cf1953}. 

As the density in the core increases flux freezing breaks down, and the
relative drifts of different charged species with respect to the neutral
particles delineate three magnetic diffusivity regimes: 
\begin{itemize}
{\item the \textit{Ohmic (resistive) diffusion} limit, which dominates in high
density regions where the ionisation fraction is low. The ions and the
electrons frequently collide with the neutrals over the electron gyration
period, and the magnetic field is decoupled from all charged particles. Ohmic
diffusion is important in the innermost regions of the protostellar disc where
the density and collisional rates are high \citep[e.g.][]{sglc2006, mim2008}.} 
{\item the \textit{ambipolar diffusion} limit, which dominates in regions of
relatively low density where the fractional ionisation is high, causing the
ionised component to drift with the field through the neutrals. Ambipolar
diffusion is dominant in molecular clouds \citep{w2007}, in protostellar discs
at radial distances beyond $\sim 10$ au and close to the surface of these
discs nearer to the protostar \citep{s2009}.}  
{\item the \textit{Hall diffusion} limit, which dominates in the intermediate
regimes between ambipolar and Ohmic diffusion. The more massive particles such
as ions and charged dust grains are decoupled from the magnetic field and are
instead collisionally-coupled to the neutral gas. Hall diffusion is expected to
dominate in many regions of molecular clouds as they undergo gravitational 
collapse \citep{w2004a}, and in protostellar discs \citep{ss2002a, ss2002b}.}
\end{itemize}
In the Hall limit the magnetic response of the disc is not invariant under a
global reversal of the magnetic field \citep{wn1999} as Hall diffusion twists
the field and changes the angular momentum of the neutral fluid; in a
rotationally-supported disc this causes the gas to fall inwards if it loses
angular momentum, or outwards if it gains it. Ambipolar and Ohmic diffusion,
however, always cause the field to move in the radial direction against the
flow of the neutrals -- reversing the direction of the field does not affect
the direction of the field diffusion. 

The actual magnitude and type of coupling that occurs between the fluid and
magnetic field in molecular clouds and protostellar discs is uncertain due to
the difficulty in obtaining detailed observations in these regions,
particularly of the magnetic field. Calculations of the ionisation equilibrium
and resistivity by \citet{w2004a} suggested that Hall diffusion is important
and may dominate the magnetic field behaviour at many of the densities and
field strengths encountered in molecular clouds and protostellar discs. In
particular, the Hall term is significant for molecular gas densities in the
range $\sim 10^8$--$10^{11}$ cm$^{-3}$ (when $B$ scales as $B \propto
n_\text{H}^{1/4}$), although the presence and distribution of grains
complicates the calculation of the diffusivities \citep{wn1999}. 

In simulations of star formation where the magnetic field behaviour is not
governed by IMHD it is usually ambipolar diffusion that is included
\citep[e.g.][]{ck1998, as2007, ml2009}. Ohmic diffusion is important in the
innermost regions of the collapse as the density builds up, particularly when
$\rho > 10^{11}$ cm$^{-3}$ \citep{sglc2006}. It seems likely that star
formation requires all of these processes to some degree, however it is only
recently that simulations have been performed with more than one of these
processes included \citep{lks2011}, and Hall diffusion is almost always
overlooked. The nature of the coupling determines the magnetic field direction
as the field lines emerge from the surface of the protostellar disc, which in
turn controls the amount of material that is able to slide along the field
lines and be flung outwards from the surface in disc-driven wind models
\citep{wk1993, w2004b}. 

In this paper we construct a semianalytic model of gravitational collapse in
order to demonstrate the importance of Hall diffusion and its influence on the
magnetic braking catastrophe, in which the magnetic braking affecting a
collapsing flow is such that all angular momentum is removed from the flow and
no Keplerian disc may form. The paper is organised as follows: in \S2 we
describe the formulation of the self-similar collapse equations, the
assumptions governing this approximation and the boundary conditions of the
core; the numerical procedures are outlined in \S3; in \S4 we present the
similarity solutions; the effect of the Hall term on the solutions and the
magnetic braking catastrophe is discussed in \S5 and our conclusions are
stated in \S6. 

\section[]{Formulation}\label{equations}

The goal of this work is to construct a semianalytic model of gravitational
collapse similar to that of KK02, including terms for Hall diffusion in the
equations for the magnetic field diffusion and braking. This allows the
calculation of similarity solutions that show the importance of Hall diffusion
in molecular cloud cores and collapsing flows, as well as comparisons between
the influence of the Hall and ambipolar diffusion terms. Following KK02, the
magnetohydrodynamic equations for the isothermal system are given by 
\begin{equation}
    \frac{\partial\rho}{\partial{t}} + 
    \nabla\cdot(\rho\mathbf{V}) = 0, \label{m1}
\end{equation}
\begin{equation}
    \rho\frac{\partial\mathbf{V}}{\partial{t}}
    + \rho(\mathbf{V}\cdot\nabla)\mathbf{V} = -\nabla{P} + \rho\mathbf{g}
    + \mathbf{J}\times\mathbf{B}, \label{rm1}
\end{equation}
\begin{equation}
    \nabla^2\Phi = 4\pi G\rho, \label{pot1}
\end{equation}
\begin{equation}
    \nabla\cdot\mathbf{B} = 0, \label{am1}
\end{equation}
and
\begin{align}
    \frac{\partial{\mathbf{B}}}{\partial{t}} &= \nabla \times (\mathbf{V} 
    \times \mathbf{B}) \nonumber \\ &-\!\nabla\!\times\! 
    \left[\eta\!\left(\nabla\!\times\!\mathbf{B}\right) + 
    \eta_{H}\!\left(\nabla\!\times\!\mathbf{B}\right)\!\times\!\mathbf{\hat{B}} 
    + \eta_{A}\!\left(\nabla\!\times\!\mathbf{B}\right)_{\perp}\!\right]\!,
    \label{in1}
\end{align}
where $\rho$ is the gas density, $\mathbf{V}$ the velocity field, $P$ the
pressure, $\mathbf{g}$ the gravitational field, $\Phi$ the gravitational
potential, $c$ the speed of light, $\mathbf{J}$ the current density,
$\mathbf{B}$ the magnetic field, and $\eta$ and $\eta_{H,A}$ are the
diffusion coefficients for the Ohmic, Hall and ambipolar terms in the
induction equation. 

We assume that the collapse is axisymmetric to simplify the calculations, and
the magnetic field is taken to be aligned with the axis of rotation. This is
in contrast with some observations of protostellar systems such as the binary
NGC 1333 IRAS 4A where the axis normal to the binary envelope lies between the
outflow and magnetic field axes \citep{grm2006, aetal2009}, however most
observations seem to show that the magnetic field takes on an hourglass shape
that is aligned with the axis of rotation \citep[e.g.][]{vetal2005, cc2006}.
Alignment is required by the assumption of axisymmetry, as misalignment
introduces three-dimensional effects that cannot be modelled here. 

Using cylindrical coordinates, the mass, radial momentum and angular momentum
conservation equations, as well as the hydrostatic equilibrium equation, the
solenoidal condition and the vertical component of the induction equation are,
under the assumptions of isothermality (that $P = \rho c_s^2$) and
axisymmetry: 
\begin{equation}
    \frac{\partial\rho}{\partial{t}} 
      + \frac{1}{r}\frac{\partial}{\partial{r}}(r\rho{V_{r}}) 
      = - \frac{\partial}{\partial{z}}(\rho{V_{z}}),\label{m2}
\end{equation}
\begin{align}
    \rho\frac{\partial{V_{r}}}{\partial{t}} 
      &+ \rho{V_{r}}\frac{\partial{V_{r}}}{\partial{r}} 
      = \rho{g_{r}} - c_s^{2}\frac{\partial{\rho}}{\partial{r}} 
      + \rho\frac{V^{2}_{\phi}}{r}
      + \frac{B_{z}}{4\pi}\frac{\partial{B_{r}}}{\partial{z}} 
      \nonumber \\
      &\quad- \frac{\partial}{\partial{r}}\!\!\left(\!\frac{B_{z}^{2}}{8\pi}\!\right) 
      - \frac{1}{8\pi{r^{2}}}\frac{\partial}{\partial{r}}(rB_{\phi})^{2} 
      - \rho{V_{z}}\frac{\partial{V_{r}}}{\partial{z}}, \label{rm2}
\end{align}
\begin{align}
    \frac{\rho}{r}\frac{\partial}{\partial{t}}(rV_{\phi}) 
      &+ \frac{\rho{V_{r}}}{r}\frac{\partial}{\partial{r}}(rV_{\phi}) \nonumber\\ 
      &= \frac{B_{z}}{4\pi}\frac{\partial{B_{\phi}}}{\partial{z}} 
      + \frac{B_{r}}{4\pi{r}}\frac{\partial}{\partial{r}}(rB_{\phi})
      - \rho{V_{z}}\frac{\partial}{\partial{z}}(r{V_{\phi}}), 
      \label{am2}
\end{align}
\begin{align}
    \rho\frac{\partial V_z}{\partial t} 
      + \rho V_r\frac{\partial V_z}{\partial r}
      &+ \rho V_z\frac{\partial V_z}{\partial z}
      + c_s^{2}\frac{\partial{\rho}}{\partial{z}} \nonumber\\
      & = \rho{g_{z}} 
      - \frac{\partial}{\partial{z}}\!\left(\!\frac{B^{2}_{\phi}}{8\pi} 
      + \frac{B^{2}_{r}}{8\pi}\!\right)\! 
      + \frac{B_{r}}{4\pi}\frac{\partial{B_{z}}}{\partial{r}},
      \label{he2}
\end{align}
\begin{equation}
    \frac{\partial{B_z}}{\partial{z}} = 
      -\frac{1}{r}\frac{\partial}{\partial{r}}(rB_r) \label{solenoid}
\end{equation}
and
\begin{align} 
    \frac{\partial{B_{z}}}{\partial{t}} 
    = -\frac{1}{r}\frac{\partial}{\partial{r}} \Biggl[r \Bigl(\!V_{r}B_{z} 
      &+ \Big[\eta (\nabla \!\times\! \mathbf{B})
      + \frac{\eta_{H}}{B}(\nabla \!\times\! \mathbf{B}) \!\times\! \mathbf{B}
      \nonumber\\
    &- \frac{\eta_{A}}{B^{2}}\left((\nabla \!\times\! \mathbf{B}) \!\times\! 
      \mathbf{B}\right) \!\times\! \mathbf{B}\Big]_{\phi}\Bigr)\!\Biggr]
      \label{in2}
\end{align}
where $g_{r}$ and $g_{z}$ are the radial and vertical components of the
gravitational field, and $c_s$ is the isothermal sound speed given by $c_s =
(k_BT/m_n)^{1/2} \approx 0.19$ km s$^{-1}$ (with $k_B$ the Boltzmann constant,
$T$ the gas temperature, typically taken to be $10$ K, and $m_n$ the mean mass
of a gas particle). The assumption of isothermality breaks down due to
radiative trapping when the central density reaches $\sim 10^{10}$ cm$^{-3}$
\citep{g1963}, which occurs on scales $r \lesssim 5$ au for a typical
simulation. It is expected that isothermality shall break down in the
innermost regions of our solutions, however, as thermal stresses do not play a
significant role in the larger-scale dynamics isothermality is not expected to
introduce large errors into the calculations. 

The disc is assumed to be thin based upon the results of Mouschovias and
collaborators \citep[e.g.][]{fm1992, fm1993}, which have shown that an
initially-uniform, self-gravitating, magnetised molecular cloud core rapidly
collapses along the magnetic field lines. This assumption allows us to further
reduce the dimensionality of the problem by vertically-averaging the variables
over the scale height of the disc, although it implies that processes that
depend on variations in the density or the magnetic field with height within
the disc cannot be included in the collapse calculations. Effects such as
turbulence or the interaction between active and dead zones in the disc are
not expected to have a large effect on the overall dynamics of early collapse
(although they are known to become important in some regions of protostellar
discs once the adiabatic core and protostar have formed) and their exclusion
is necessary to the self-similarity of the solutions. 

The vertical averaging is performed as in KK02. As only the vertical component
of the induction equation and our equation for the magnetic braking differ
from theirs, we perform the integration of these in Appendix \ref{vertavg} and
refer the reader to their appendix A for the others. The quantities
$\eta$, $\eta_H/B$ and $\eta_A/B^2$ are approximated as being constant with
height, as are the radial velocity, the azimuthal velocity and the radial
component of gravity. 

The disc is threaded by an open magnetic field possessing an even symmetry,
so that $B_r = B_\phi = 0$ at the disc midplane. The radial and toroidal field
components are taken to scale with height as 
\begin{align}
    B_r(r,z) &= B_{r,s}(r)\frac{z}{H(r)} \label{Brscale}\\
    B_\phi(r,z) &= B_{\phi,s}(r)\frac{z}{H(r)} \label{Bphiscale}
\end{align}
where $H(r)$ is the scale height of the disc and $B_{r,s}$ and $B_{\phi,s}$
are the magnetic field components at the surface of the disc; these scalings
are motivated by the field configuration of a rotationally-supported thin disc
in which the field is well-coupled to the gas \citep{wk1993}. As ambipolar
diffusion and Hall diffusion become more important in the inner regions of the
disc where the field is less well-coupled to the gas, this approximation is no
longer adequate, however none of the dominant terms in the equation set depend
upon the particulars of the vertical variation of the field within the disc
(KK02), so it remains reasonable to adopt these scalings across the domain of
self-similar collapse. 

The surface density of the pseudodisc is defined by the expression
\begin{equation}
    \Sigma = \int^\infty_{-\infty} \rho dz = 2H\rho, \label{Sigdef}
\end{equation}
assuming that the density is constant with height within the disc, and the
specific angular momentum is defined by $J = rV_\phi$. We neglect any mass
loss due to a disc wind. The radial components of gravity and the magnetic
field are calculated using the monopole expressions 
\begin{equation}
    g_r = -\frac{GM(r)}{r^2} \label{gr}
\end{equation}
and 
\begin{equation}
    B_{r,s} = \frac{\Psi(r,t)}{2\pi{}r^2}, \label{Brs}
\end{equation}
where the enclosed mass $M(r) \approx M_c$ when the central mass dominates and
$\Psi$ is the magnetic flux enclosed within the radius $r$. These
simplifications were also used by KK02, as \citet{cck1998} found that these
expressions give values of the gravitational force and magnetic field that are
near enough to those found using an iterative method that they do not
introduce significant errors into the calculation. 

The vertical angular momentum transport above and within the disc is achieved
by magnetic braking, especially during the dynamic collapse phase inwards of
the magnetic diffusion shock and in the rotationally-supported inner disc. The
approach to modelling the magnetic braking used in this work, which is
described in more detail in Appendix \ref{vertavg}, is adapted from that of
\citet{bm1994} for the pre-point mass formation collapse phase. This
formulation is not well-defined in the innermost rotationally-supported
regions of the pseudodisc, where the calculated magnetic braking becomes
stronger than is expected and the angular momentum transport is expected to be
dominated by a disc wind (which is not included but discussed further in
\S\ref{discuss}).

A cap is then placed upon the azimuthal magnetic field component in order to
ensure that it does not greatly exceed the vertical component; balancing the
torques on the disc then allows us to define 
\begin{equation}
    B_{\phi,s} = -\mathrm{min}\left[\frac{\Psi}{\pi{r^2}}
    \left(\frac{r\Omega - r\Omega_b}{V_\text{A,ext}}\right);\delta{B_z}\right];
\label{b_phis1}
\end{equation}
where $\delta$ is a constant parameter ($=1$ in our solutions) that limits the
magnetic braking, $\Omega$ is the local angular velocity, and $\Omega_b$ is
the background angular velocity of the cloud, which is small compared to that
of the core. $V_\text{A,ext}$ is the Alfv\'en wave speed in the external
medium, which is parameterised with respect to the sound speed using 
\begin{equation}
    \alpha = c_s/V_\text{A,ext},\label{alpha}
\end{equation}
where again $\alpha$ is a constant parameter of the model. We take $\alpha =
0.08$ in the solutions presented here, which is a reasonable approximation as
the observations of \citet{c1999} indicated that $V_\text{A,ext} \approx 1$ km
s$^{-1}$ over many orders of magnitude in several molecular clouds. The
negative sign of $B_{\phi,s}$ and positive sign of $\delta$ are based on the
assumption that there is no counter-rotation of material in the collapse; were
the fluid counter-rotating then the signs of $\delta$ and $B_{\phi,s}$ would
both change. The azimuthal drift velocity of the magnetic field is averaged
over the disc scale height in Appendix \ref{vertavg} to give the final form of
Equation \ref{b_phis} shown below. 

The disc equations are further simplified by recognising that the thin disc
approximation implies that terms of order $\mathcal{O}(H/r)$ are small in
comparison to the other terms and can then be dropped from the equations. As
in KK02, the only term of order $\mathcal{O}(H/r)$ that is kept is the
combination $[B_{r,s} - H(\partial B_z/\partial z)]$, which occurs in the
radial momentum equation and is important in refining the structure of the
magnetic diffusion shock. This term is then retained in all of the equations
in which it appears. 

Taking all of these into account then gives the simplified set of equations: 
\begin{align}
    &\frac{\partial\Sigma}{\partial{t}} + 
	\frac{1}{r}\frac{\partial}{\partial{r}}(r\Sigma{V_{r}}) = 0,
	\label{mass1}\\ \nonumber\\
    &\frac{\partial{V_{r}}}{\partial{t}} + {V_{r}}\frac{\partial{V_{r}}}{\partial{r}} 
    	= g_r - \frac{c_s^{2}}{\Sigma}\frac{\partial\Sigma}{\partial{r}} 
    	+ \frac{B_{z}B_{r,s}}{2\pi\Sigma} + \frac{J^2}{r^3}, \label{rad1}\\
 	\nonumber\\
    &\frac{\partial{J}}{\partial{t}} + V_{r}\frac{\partial{J}}{\partial{r}}
    	= \frac{rB_{z}B_{\phi,s}}{2\pi\Sigma},\label{ang1}\\ \nonumber\\
    &\frac{{\Sigma}c_s^2}{2H} 
        = \frac{\pi}{2}G\Sigma^2 + \frac{GM_{c}\Sigma{H}}{4r^3}
        + \frac{1}{8\pi}\left(B_{r,s}^{2} + B_{\phi,s}^2\right),\label{vert1}\\
	\nonumber\\
    &\frac{H}{2\pi}\frac{\partial\Psi}{\partial{t}}
        = -rHV_rB_z - \eta{B_{r,s}}
        - \frac{r\eta_H}{B}B_zB_{\phi,s}
        - \frac{r\eta_A}{B^2}B_{r,s}B_z^2 \label{ind1}
\end{align}
and 
\begin{align}
    B_{\phi,s} = -\min\Biggl[\frac{\Psi\alpha}{\pi{r^2}c_s}
        &\left[\frac{J}{r} - \frac{\eta_H}{B}\left(\!B_{r,s} 
        - H\frac{\partial{B_z}}{\partial{r}}\!\right)\!\right]\nonumber \\
    &\left[1 + \frac{\Psi\alpha}{\pi{r^2}c_s}\frac{\eta_P}{B^2}
       \frac{B_z}{H}\right]^{-1};\delta{B_z}\Biggr],
\label{b_phis2}
\end{align}
along with Equations \ref{gr} and \ref{Brs}.

\subsection{Self-Similarity}\label{ssim}

At any instant in time the collapse solutions look like stretched versions of
themselves at previous times; this fractal-like behaviour is referred to as
self-similarity. The pseudodisc forms as a collapse wave (referred to as the
magnetic diffusion shock) propagates outwards at the speed of sound. The
self-similarity of the waves of infall occurs because of the lack of
characteristic time and length scales in the flow. 

Gravitational collapse occurs over many orders of magnitude in radius and
density, so that the point mass has negligible dimensions in comparison with
the accretion flow.  The only dimensional quantities that effect the flow are
the magnetic field $\mathbf{B}$, the diffusion coefficients $\eta$ and
$\eta_{H, A}$, the gravitational constant $G$, the isothermal sound speed
$c_s$, the local radius $r$ and the instantaneous time $t$; this means that,
except for scaling factors, all of the flow variables may be written as
functions of a similarity variable defined by 
\begin{equation}
    x = \frac{r}{c_st}.
  \label{x}
\end{equation}
KK02 noted that for a typical value of the sound speed ($c_s = 0.19$ km
s$^{-1}$ at $T = 10$ K), $x = 1$ corresponds to a distance of $r \approx 6
\times 10^{15}$ cm (400 au) when $t=10^4$ yr (which is the characteristic age
of a Class 0 YSO) and to a distance of $r\approx6\times10^{16}$ cm (4,000 au)
when $t=10^5$ yr (the age of a Class 1 YSO). The Class 0 YSO IRAM 04191 has a
dense inner disc-like structure that resembles a tilted ring with an average
radius of $r_0 \sim 1400$ au \citep{lhw2005} --- this is of the same order of
magnitude as the centrifugal shock radius in the disc-forming solutions at the
same age.

The physical quantities are expressed as the product of a nondimensional flow
variable that depends only upon $x$ and a dimensional part constructed from
$c_s$, $G$ and $t$: 
\begin{align}
    \Sigma(r,t) &= \left(\frac{c_s}{2\pi{Gt}}\right)\sigma(x), \label{sssig}\\
       g_r(r,t) &= \left(\frac{c_s}{t}\right)g(x),\label{ssG}\\
    V_r(r,t) &= c_su(x), \label{ssvr}\\
      H(r,t) &= c_sth(x), \label{ssh}\\
    V_\phi(r,t) &= c_sv(x), \label{ssvphi}\\
         J(r,t) &= c_s^2tj(x), \label{ssj}\\
    M(r,t) &= \left(\frac{c_s^3t}{G}\right)m(x), \label{ssm}\\
    \dot{M}(r,t)&= \left(\frac{c_s^3}{G}\right)\dot{m}(x),\label{ssMdot}\\ 
    \mathbf{B}(r,t) &= \left(\frac{c_s}{G^{1/2}t}\right)\mathbf{b}(x),\label{ssb}\\
         \Psi(r,t) &= \left(\frac{2\pi{c_s^3t}}{G^{1/2}}\right)\psi(x),\label{sspsi}\\ 
    \text{and }\eta_{H,A} &= c_s^2t\eta'_{H,A}. \label{sseta} 
\end{align}
These equations have the same form and use the same notation as those in KK02,
with the addition of extra diffusion coefficients to model the magnetic field
more completely. 

The Ohmic and ambipolar diffusion terms scale together, to a zeroth-order
approximation, as they possess a similar dependence upon $B$ and appear in the
induction equation and the equation for the azimuthal field component
multiplied by the same magnetic field terms. Because the field within the thin
disc is effectively vertical, both ambipolar and Ohmic diffusion influence the
field drift in the same manner. While one type of diffusion may dominate over
the other at any individual point in the disc \citep[in general, ambipolar
diffusion in the outer regions where the density is low and Ohmic diffusion in
the inner regions where the density is high;][]{w2007}, only one term is
needed in order to study the change in the disc behaviour introduced by the
Hall diffusion term that is of most interest in this work. The Ohmic and
ambipolar diffusion terms are combined into a single term parameterised by the
dimensionless constant $\tilde{\eta}_A$, referred to as the ambipolar
diffusion parameter. 

The ambipolar diffusion coefficient in a molecular cloud core without grains
is given by the equation 
\begin{equation}
    \eta_A = \frac{B^2}{4\pi\gamma\rho_i\rho},
\label{etaa}
\end{equation}
where $1/\gamma\rho_i = \tau_{ni}$ is the neutral-ion momentum exchange
timescale, parameterised as
\begin{equation}
    \tau_{ni} = \frac{\tilde{\eta}_A}{\sqrt{4{\pi}G\rho}};
\label{tauni}
\end{equation}
the nondimensional ambipolar diffusion parameter $\tilde{\eta}_A$ is a constant
of the model (simply denoted $\eta$ in KK02). $\eta_A/B^2$ is then
self-similarised using the scalings above to give 
\begin{equation}
    \frac{\eta_A'}{b^2} = \tilde{\eta}_A\frac{h^{3/2}}{\sigma^{3/2}};
\label{ssetaa}
\end{equation}
it is important to note that the self-similarity of the solution depends upon
the relationship $\rho_i \propto \rho^{1/2}$. 

For grains with radius $a = 0.1$ $\mu$m in a cloud where the temperature is
$10$ K and the cosmic ray ionisation rate is $\xi = 10^{-17}$ s$^{-1}$,
simulations typically assume the ion density scales as $\rho_i \propto
\rho_n^{1/2}$ when $10^4 \lesssim n_\text{H} \lesssim 10^7$ cm$^{-3}$
\citep{e1979, kn2000}. This is an oversimplification, as \citet{cm1998} showed
that for typical cloud and grain parameters the proportionality of the ion
density cannot be parameterised by a single power law exponent, but it is
still a reasonable and widely-adopted approximation to the ion density in
collapsing cores on scales $\gtrsim 10^3$ au \citep[see e.g.][KK02]{sal1987,
gs1993a, ck1998, cck1998}. 

As a matter of pragmatism, a similar scaling with respect to the density and
scale height is adopted for the Hall diffusion parameter, $\eta_H$. By stating
that the self-similar Hall diffusion coefficient scales as 
\begin{equation}
    \frac{\eta'_H}{b} = \tilde{\eta}_Hb\frac{h^{3/2}}{\sigma^{3/2}}
\label{ssetah}
\end{equation}
where $\tilde{\eta}_H$ is the constant nondimensional Hall diffusion parameter
used to characterise the solutions, the ratio of the nondimensional ambipolar
and Hall diffusion parameters becomes the most important factor in determining
the magnetic behaviour of the similarity solutions. In truth, the Hall
diffusion coefficient could be scaled with respect to the density and field
strength by multiplying the nondimensional Hall parameter by any function of
the similarity variable $x$ and the fluid variables. This topic is discussed
in more detail in \S \ref{discuss}, where an alternate scaling is proposed
for future work on the self-similar collapse model. The scaling of $\eta'_H$
given in Equation \ref{ssetah} is appropriate for a molecular cloud core with
grains acting as the dominant positive charge characters. 

For convenience the variable $w\equiv{}x-u$ is used to simplify the equation
set. The similarity variables are then used to rewrite Equations
\ref{mass1}--\ref{ind1} in self-similar form: 
\begin{align}
   \frac{d\psi}{dx} &= xb_z, \label{sscpsi} \\
   \frac{dm}{dx} &= x\sigma, \label{sscm} \\
   (1-w^2)\frac{1}{\sigma}\frac{d\sigma}{dx} &= g 
	+ \frac{b_z}{\sigma}\left(\!b_{r,s} - h\frac{db_z}{dz}\!\right)
   	+ \frac{j^2}{x^3} + \frac{w^2}{x}, \label{sscrm} \\
   \frac{dj}{dx} &= \frac{1}{w}\left(j 
   	- \frac{xb_zb_{\phi,s}}{\sigma}\right), \label{sscam} \\
   \biggl(\!\frac{\sigma{m_c}}{x^3} - b_{r,s}\frac{db_z}{dx}\!&\biggr)h^2
   	+ \left(b_{r,s}^2 + b_{\phi,s}^2 + \sigma^{2}\right)h 
	- 2\sigma = 0, \label{ssvhe} 
\end{align}
and
\begin{align}
   \psi - xwb_z &+ \tilde{\eta}_Hxb_{\phi,s}b_zbh^{1/2}\sigma^{-3/2}
	\nonumber \\
   &+ \tilde{\eta}_Axb_z^2h^{1/2}\sigma^{-3/2}\left(\!b_{r,s} 
      - h\frac{db_z}{dx}\!\right) = 0. \label{ssin}
\end{align}
These equations are augmented by the self-similar definitions 
\begin{align}
   m &= xw\sigma,\label{ssmxw}\\
   \dot{m} &= -xu\sigma, \label{ssmdot} \\
   \text{and }g &= -\frac{m}{x^2}; \label{ssg}
\end{align}
while the other magnetic field components are given by 
\begin{equation}
   b_{r,s} = \frac{\psi}{x^2}
\label{ssb_r}
\end{equation}
and
\begin{align}
   b_{\phi,s} = -\min\Biggl[\frac{2\alpha\psi}{x^2} &\left[\frac{j}{x} 
    -\frac{\tilde{\eta}_Hh^{1/2}b}{\sigma^{3/2}}\!\left(\!b_{r,s} 
    -h\frac{db_z}{dx}\!\right)\right] \nonumber \\
   &\left[1 + \frac{2\alpha\tilde{\eta}_Ah^{1/2}\psi{b_z}}{x^2\sigma^{3/2}}
    \right]^{-1};\delta{b_z}\Biggr]. \label{ssb_phis}
\end{align}
These equations completely describe the collapse of the molecular cloud core
into a pseudodisc and the accretion onto the central point mass (potentially
through a rotationally-supported disc). These equations are the same as
equations 20--32 of KK02 in the limit of $\tilde{\eta}_H = 0$, which allows
direct comparisons to be made between the similarity solutions of both models.

\subsection{Outer boundary conditions}\label{outer}

The outer regions of the collapse are modelled by a set of power law relations
in the similarity variable that describe a molecular cloud core contracting
quasistatically under ambipolar diffusion until it has just become
supercritical and a point mass forms at the centre
\citep[e.g.][]{s1977,sal1987}. 

The definition of the similarity variable (Equation \ref{x}) means that the
limit $x \to \infty$ corresponds both to the outer edge of the core at $r \to
\infty$ and the initial conditions of the collapse as $t \to 0$. We then
describe the core as a singular isothermal sphere, which has the density
profile $\rho \propto r^{-2}$ \citep[e.g.][]{l1969,p1969,ws1985}, so that the
surface density is: 
\begin{equation}
    \Sigma (t = 0) = \frac{Ac_s^2}{2\pi Gr} \label{outSig}
\end{equation}
where $A$ is a constant determined by the initial accretion rate of the core.
The infall velocity and accretion rate onto the core are constant and given by
\begin{align}
    V_r (t = 0) &= u_0c_s \label{outVr}\\
    \text{and } \dot{M} (t = 0) &= -\frac{Au_0c_s^3}{G}. \label{outM}
\end{align}
The numerical results of \citet{ck1998} for the collapse of a rotating
magnetic core with ambipolar diffusion showed that the accretion rate at point
mass formation was $\dot{M} \simeq 5$ M$_\odot$ Myr$^{-1}$, corresponding to a
nondimensional parameter $A \simeq 3$, which also matches to observations of
many cores that show $\dot{M} \in [1,10]$ M$_\odot$ Myr$^{-1}$
\citep{lmt2001}. As the isothermal sound speed of a core at $T = 10$ K is
equal to $c_s = 0.19$ km s$^{-1}$, the nondimensional initial infall speed
must be of order unity to match observations showing $V_r = 0.05$--$0.10$ km
s$^{-1}$ \citep{lmt2001}; to match the parameter of KK02 we adopt $u_0 = -1$. 

The rotational velocity of the initial core is spatially-uniform and given by 
\begin{equation}
    V_\phi = v_0c_s \label{outVphi}
\end{equation}
where $v_0$ is the dimensionless rotational velocity, which can be
approximated by 
\begin{equation}
    v_0 \approx \frac{A\Omega_bc_s}{\sqrt{G}B_\text{ref}} \label{outv0}
\end{equation}
using the $r^{-1}$ dependence of the core surface density and magnetic field
as in \citet{b1997}. The uniform background angular velocity is typically
$\Omega_b = 2 \times 10^{-14}$ rad s$^{-1}$ \citep{gbfm1993, kc1997} and the
background magnetic field is taken to be $B_\text{ref}= 30$ $\mu$G
\citep{c1999}, which gives $v_0 = 0.15$. This value is a factor of ten larger
than that obtained in \citet{b1997}, however, the range of observed core
velocities is $v_0 \in [0.01, 1.0]$ \citep{lmt2001}. KK02 showed that the size
of the inner Keplerian disc directly corresponds to the initial rotational
velocity at the core; we take $v_0 = 0.73$ to facilitate comparison with their
fiducial ambipolar diffusion solution. 

Finally, in the outer regions IMHD holds true, and the mass-to-flux ratio in
the gas is constant. The core is just supercritical at 
\begin{equation}
    \frac{M}{\Psi} = \frac{\mu_0}{2\pi\sqrt{G}} \label{outMPsi}
\end{equation}
where $\mu_0$ is the dimensionless mass-to-flux ratio, parameterised with
respect to the critical value for support against gravity \citep{nn1978}. A
value of $\mu_0 = 2.9$ is adopted here for compatibility with KK02; this value
was obtained from the numerical simulations of \citet{ck1998} and matches
observations showing that cores are typically more than twice supercritical
\citep[Crutcher 1999; but see also][]{cht2009, mt2009}. Equation \ref{sscpsi}
can then by used to show that 
\begin{equation}
    B_z = \frac{2\pi\sqrt{G}}{\mu_0}\Sigma; \label{outBz}
\end{equation}
this is equivalent to Equation \ref{outMPsi}, so that only one of these may be
used to calculate the similarity solution.  

In self-similar form the conditions at the outer boundary $x_\text{out}$ take the
form
\begin{align}
    m &= Ax_\text{out}, \label{outm2}\\
    \sigma &= \frac{A}{x_\text{out}}, \label{outsigma2}\\
    b_z &= \frac{\sigma}{\mu_0} \label{outbz2}\\
    \text{and } v &= v_0. \label{outv2}
\end{align}

\subsection{Inner boundary conditions}\label{innerbc}

As the variables are integrated inward from the centrifugal shock they tend
towards an inner asymptotic set of equations describing a Keplerian disc
around the protostar. These relations are found analytically by assuming that
the variables take the form of power laws in $x$ and solving Equations
\ref{sscpsi}--\ref{ssb_phis} for the exponents as $x \to 0$, under the
assumption that there is no counter-rotation. In nondimensional form this
disc is described by: 
\begin{align}
    m &= m_c, \label{in-m}\\  
    \dot{m}&= m_c, \label{in-mdot}\\
    \sigma &= \sigma_1\,x^{-3/2} 
     = \frac{\sqrt{2m_c}f}{2\delta\sqrt{(2\delta/f)^2 + 1}}\,x^{-3/2}, 
     \label{in-sigma}\\
    h &= h_1\,x^{3/2} 
     = \left(\frac{2}{m_c[1+(f/2\delta)^2]}\right)^{\!1/2}x^{3/2}, \label{in-h}\\
    u &= -\frac{m_c}{\sigma_1}\,x^{1/2}, \label{in-u}\\
    v &= \sqrt{\frac{m_c}{x}}, \label{in-v}\\
    j &= \sqrt{m_cx}, \label{in-j}\\
    \psi &= \frac{4}{3}b_zx^2, \label{in-psi}\\
    b_z &= \frac{m_c^{3/4}}{\sqrt{2\delta}}\,x^{-5/4}, \label{in-bz}\\
    b_{r,s} &= \frac{4}{3}\,b_z \label{in-brs}\\
    \text{and } b_{\phi,s} &= -\delta\,{b_z}. \label{in-bphis}
\end{align}
The diffusion constant $f$ is a function of the ambipolar and Hall diffusion
parameters, and is given by the equation 
\begin{equation}
    f = \frac{4}{3}\,\tilde{\eta}_A 
      - \delta\tilde{\eta}_H\sqrt{\frac{25}{9} + \delta^2}; \label{in-f}
\end{equation}
this definition shows how the Hall term is able to counteract the ambipolar
diffusion term in determining the surface density of the disc and the
accretion rate onto the central protostar when the nondimensional Hall
parameter $\tilde{\eta}_H$ is positive, and add to the ambipolar diffusion if
the Hall parameter is negative. The characteristic diffusion parameter of the
disc, $f$, must be positive in order to form a Keplerian disc in solutions
without counter-rotation\footnote{As discussed in \S5, the Hall effect can
cause counter-rotation of the disc; however, in all of our solutions the inner
Keplerian disc is rotating in the same direction as the initial core.}; this 
places limits on the relative sizes of the diffusion parameters.

These relations are derived elsewhere (Braiding \& Wardle, in prep): to
summarise, the inner accretion disc is in Keplerian rotation in the same
direction as the initial core, with the centrifugal force balancing the
inwards pull of gravity. The accretion rate onto the protostar is constant and
low, and accretion through the disc is determined by the total amount of
magnetic diffusion, which removes radial support by the magnetic field and
allows the gas to fall inwards. 

A second asymptotic solution also exists (KK02; Braiding \& Wardle, in prep),
in which the magnetic braking is so strong that most of the angular momentum
is removed from the gas, which then free falls onto the protostar without
forming a disc (the magnetic braking catastrophe, described in
\S\ref{discuss}). Similarity solutions matching onto this solution lie beyond
the scope of the present paper. 

\section{Numerical Methods}\label{numerics}

The calculation of the similarity solutions is quite complex, as many of the
derivatives are large, and sonic points and shock fronts must be calculated
explicitly as they are encountered by the integration routine. Furthermore,
subshocks can occur downstream of the magnetic diffusion and centrifugal
shocks. Such a subshock occurred downstream of the magnetic diffusion shock in
the solutions of \citet{l1998}, and although no subshocks appeared in the
published solutions of KK02 they were observed in the unpublished collapse
solutions discussed in that paper. Locating and integrating through
these shocks requires careful monitoring of the integration and automatic
intervention where necessary. The full details of the methods used to compute
the similarity solutions are described in \S5.1 of \citet{b2011}; only an
outline is provided in this work.  

The problem is recast as a two-point boundary value problem in which the
variables are integrated from a matching point $x_m$ to both the inner and
outer boundaries. We employ a ``shooting'' routine, which modifies the values
of the variables at $x_m$ in order to zero the discrepancies between the
integrated variables and their expected asymptotic values at the boundaries.
The matching point is located at a position $x_c < x_m < x_d$, where $x_d$ and
$x_c$ are the positions of the magnetic diffusion and centrifugal shocks
respectively, and is typically chosen just downstream of the magnetic
diffusion shock at $x_m \sim 0.3$. 

The values of the variables $m$, $\sigma$, $j$, $\psi$ and $b_z$ at $x_m$ and
the nondimensional accretion rate onto the central mass $m_c$ are initially
unknown, and a poor guess of these can cause the integration (and by extension
convergence on the true solution) to fail. An acceptable initial guess of the
variables is obtained by performing a simplified integration from the outer
boundary to the matching point. 

\subsection{Initial guess at $x_m$}\label{num:simp}

The initial guess of the variables at the matching point is estimated by
performing a simplified integration in which the induction equation has been
replaced by an algebraic expression for the vertical magnetic field component.
We follow the derivation of KK02 here, and in the ambipolar diffusion
($\tilde{\eta}_H = 0$) limit our equations reduce to theirs. 

The initial guess of $m_c$ is estimated from the mass plateau in the region of
the magnetic diffusion shock (see \S\ref{results}). The outer edge of this
plateau occurs at $x_{pl} \approx |u_0|$ (which is something of an
overestimate as $|u|$ is typically larger than $x$ here). The outer asymptotic
behaviour is only just starting to break down in this region, and so the
asymptotic relations from \S\ref{outer} are substituted into Equation
\ref{ssmxw}: 
\begin{equation}
    m_c \approx m_{pl} \approx 2|u_0|A. \label{mcguess}
\end{equation}
The calculation of the similarity solutions is less sensitive to our estimate
of $m_c$ than to the values of the other variables at the matching point.
Equation \ref{mcguess} is an acceptable first guess of the accretion rate onto
the central star. 

Save for during the various shock transitions where $b_z$ changes rapidly, the
inequality $b_{r,s} \gg h(db_z/dx)$ holds true everywhere in the collapsing
flow; as this term is always small compared to the other terms in the
induction equation it may be dropped. The magnitude of the magnetic field $b$
is always of order $b_z \approx b_{r,s}$, so that it may be estimated by $b
\approx \sqrt{2}b_z$. The induction equation may then be written as a
quadratic in $b_z$:  
\begin{equation}
    xh^{1/2}\sigma^{-3/2}
     \left(\tilde{\eta}_H\sqrt{2}b_{\phi,s} 
    + \tilde{\eta}_Ab_{r,s}\right)b_z^2 - xwb_z + \psi = 0.
\label{hall-simpin}
\end{equation}
The well-separated roots provide approximations to the behaviour of $b_z$ on
either side of the magnetic diffusion shock, with the prescription 
\begin{equation}
    b_{z,low} \approx \frac{\psi\sigma}{m} \approx \frac{\sigma}{\mu_0}
\label{bzlow}
\end{equation}
applying in the large $x$ regime where flux freezing still mostly holds true
and the mass-to-flux ratio is given by its initial value $\mu = \mu_0$. This
is equivalent to the initial condition for the vertical field component and
although IMHD breaks down before the magnetic diffusion shock Equation
\ref{bzlow} remains a good approximation to the field in this region. 

The larger root gives the value of the vertical field component in the
magnetic diffusion regime where $x$ is small. It is approximated by dropping
the constant term in Equation \ref{hall-simpin} and solving for $b_z$ to
obtain 
\begin{equation}
    b_{z,high} \approx \frac{m}{x}\left(\frac{\sigma}{h}\right)^{1/2}
     \left(\sqrt{2}\tilde{\eta}_Hb_{\phi,s} 
     + \tilde{\eta}_Ab_{r,s}\right)^{-1},
\label{bzhigh}
\end{equation}
which reduces to equation 50 of KK02 in the $\tilde{\eta}_H = 0$ limit. The
thickness of the disc in this region is controlled by the magnetic squeezing,
so that
\begin{equation}
    h \approx \frac{2\sigma}{b_{r,s}^2}. \label{hsimp}
\end{equation}

The transition between the two approximations to $b_z$ occurs at the magnetic
diffusion shock, $x_d$, which represents a continuous increase in the magnetic
field strength, and though the matter is slowed in the post-shock region, $w$
is near-constant and $w > 1$ throughout the magnetic diffusion shock itself.
The magnetic field lines are compressed by the shock, resulting in an increase
in the vertical and azimuthal field as the field lines are twisted up by the
slowing of the compressed gas. Conservation of flux ensures that the radial
field component does not change in the shock. On the upstream side of the
magnetic diffusion shock $b_{\phi,s} \approx -\delta b_{z,low}$ (as can be
seen in the solutions in \S\ref{results}), however downstream of the shock the
azimuthal magnetic field component is given by $b_{\phi,s} \approx -w\delta
b_{z,low}$. In the region of the magnetic diffusion shock $b_{\phi,s}$ is then
simplified into 
\begin{equation}
    b_{\phi,s} \approx -x_d \delta b_{r,s}; \label{xdbphis}
\end{equation}
as $x_d < 1$, $b_{\phi,s} < b_{r,s}$ when $\delta = 1$, justifying our
omission of the azimuthal magnetic field component from the approximations to
the scale height and magnetic field amplitude. 

Then using these relations and the approximation that IMHD holds true in this
region, Equation \ref{bzhigh} becomes  
\begin{equation}
    b_{z,high} \approx \frac{m}{x} \left(\sqrt{2}\tilde{\eta}_A - 
     2\delta x_d \tilde{\eta}_H\right)^{-1};
\label{bzhighgood}
\end{equation}
applying the approximation that $b_{z,high} = b_{r,s}$ gives an estimation of
the magnetic diffusion shock position: 
\begin{equation}
    x_d \approx \frac{\tilde{\eta}_A}{\sqrt{2}}
     \left(\frac{\mu_0}{2} + \delta\tilde{\eta}_H\right)^{-1},
\label{xdgood}
\end{equation}
which is equivalent to KK02's equation $58$ in the $\tilde{\eta}_H = 0$ limit.
This value of $x_d$ is typically accurate to 20 per cent of the true value,
which is acceptable for estimating the variables at $x_m$. 

In order to obtain the initial guess of the variables at $x_m$ we integrate
Equations \ref{sscpsi}--\ref{ssvhe} inwards from the outer boundary, using
Equation \ref{bzlow} to approximate $b_z$ when $x > x_d$, and Equation
\ref{bzhigh} for $x_m < x < x_d$. These are used to integrate the full set of
equations in both directions from $x_m$; the deviation of the results from the
expected asymptotic boundary conditions is minimized by the shooting routine,
which tweaks the variables at $x_m$ to reduce the discrepancies. Once the
initial values at $x_m$ are close to the true solution the shooting routine
will converge quadratically. 

Integration of the full equation set in the outwards direction is usually
performed without difficulty unless a particularly poor guess of the variables
at $x_m$ is employed. The magnetic diffusion shock is continuous and may be
integrated through without pause. As magnetic diffusion is not important in
this region the integration to the boundary (typically located at
$x_\text{out} = 10^4$--$10^5$) is rapid and uncomplicated. Integrating in the
inwards direction is more problematic as the calculation is very sensitive
both to the values of the variables at $x_m$ and the Hall diffusion parameter. 

\subsection{Inwards integration to $x_c$}\label{num:inwards}

Integrating inwards is complicated by the presence of shock discontinuities
and sonic points, which must be calculated explicitly. Close to the magnetic
diffusion shock there may occur a subshock in which the supersonic (but
slowing due to the sudden increase in $b_z$) inflow is abruptly slowed to a
subsonic rate. This only occurs when the Hall parameter is positive, and is
likely caused by the changed magnetic braking triggered by the azimuthal field
growth during the shock. This subshock is a sharpening of the post-shock
variation of the density and infall speed in the solutions of KK02.  

The magnetic diffusion subshock's existence is detected by performing a test
integration: if the variables approach a sonic point (where $w^2 = 1$) then a
shock must exist upstream. Its position is found using the same method as that
for the centrifugal shock described below, using the same jump conditions.
Downstream of this subshock a sonic point occurs as the radial velocity
becomes supersonic once more; a small manual step through the sonic point is
performed and the variables are integrated to the centrifugal shock. 

The position of the centrifugal shock is found by performing a binary search
over an appropriate interval. The upper and lower bounds on $x_c$ are
described by $x_{c0} \pm 0.2 x_{c0}$ where $x_{c0}$ is estimated from KK02's
equation 65, using $(\tilde{\eta}_A -\delta\tilde{\eta}_H)$ in place of
$\tilde{\eta}_A$: 
\begin{equation}
    x_{c0} \approx \frac{v_0^2}{A^2}m_c 
     \exp\left[-\sqrt{\frac{2^{3/2}m_c}{\mu_0
	(\tilde{\eta}_A-\delta\tilde{\eta}_H)^3}}\right].
\label{xcbad}
\end{equation}
This is an overestimate of the true shock position (typically by around 20 per
cent), however the routine is sufficiently robust that this does not present a
problem. The variables are integrated to the estimated position of $x_{c0}$,
where the jump conditions derived below are applied, and then integrated
towards the inner boundary. Unless the shock position is known very precisely,
the variables will approach their asymptotic values and then veer off course. 

This behaviour is most clearly seen in the surface density $\sigma$, which
increases rapidly downstream of the shock if $x_{c0}$ is an overestimate to
the true shock position, and drops dramatically if $x_{c0}$ is an
underestimate. The incorrect estimate is then assigned to be the new boundary
of the shock position as appropriate, and a new estimate of $x_{c0}$ is chosen
at the midpoint between the boundaries. As the position of the shock is more
precisely known, the variables follow the asymptotic behaviour longer. 

The potential presence of any sonic points and subshocks does not interfere
with the iterative routine for finding the shock position. When there exists a
sonic point downstream of the shock: $\sigma$ bounces upwards at the sonic
point if $x_{c0}$ is too high, while if $x_{c0}$ is too low then the
integration fails at the sonic point. When the shock position is known to
approximately $10^{-10}$, it is considered to be known to the precision of the
full calculation.

The variables are integrated inwards from the centrifugal shock and if
$\tilde{\eta}_H$ is large enough they are manually integrated through the
sonic point that presages the existence of a centrifugal subshock. After the
sonic point, the binary search is employed again to find the position of its
associated shock front, as the behaviour of the variables downstream of the
subshock is unchanged. In this instance the initial upper boundary of the
search ($x_\text{up}$) is the sonic point, and the lower boundary is
$x_\text{down} = 0.1x_\text{up}$. If $\tilde{\eta}_H$ is so large that
multiple subshocks exist then this behaviour repeats, with the first subshock
followed by an additional sonic point and then a second subshock; the
increasing number of subshocks cause the calculation to become so unstable 
that it is unable to converge on the true similarity solution. 

Note that even when the shock positions are known ``precisely'', the routine
is often unable to complete the integration all the way to the inner boundary.
This is avoided by integrating a simplified set of equations from a point far
from the centrifugal shock to the inner boundary. These equations and the
matching criteria are outlined in \S\ref{num:inner}. 

\subsection{Jump conditions}\label{num:jump}

The magnetic diffusion shock is smooth and continuous, and does not require
any explicit calculation of shock conditions. The shock is the transition
between the approximations $b_{z,low}$ and $b_{z,high}$ (Equations \ref{bzlow}
and \ref{bzhigh}), and in the shock front $b_z$, $b_{\phi,s}$ and $h$ are
changed, with $b_{\phi,s}$ downstream of the shock given by $b_{\phi,s}
\approx -w x_d \delta b_{z,low}$. During the shock the scale height is
markedly compressed, suggesting that a breakdown of vertical hydrostatic
equilibrium has occurred. In reality the enhanced magnetic squeezing during
the shock front would be unable to reduce the disc thickness so dramatically
over the fluid transit time through the shock, which implies that the thinness
of the compressed region is a numerical artefact. Furthermore, as the magnetic
pressure far exceeds the gas pressure any breakdown of isothermality would not
greatly affect the collapse. 

The centrifugal shock is a true discontinuity in the fluid variables, which
must be explicitly calculated. The shock conditions used are those for the
``isothermal jump'' in KK02; these are found by recognising that the
derivatives of the surface density and radial velocity at the shock are large
compared to the other terms in the conservation of mass and radial momentum
equations (\ref{sscm} and \ref{sscrm}). The equations are integrated over the
shock to give the relations  
\begin{align}
    \sigma w &= \text{constant} \label{shock1} \\
    \text{and } \sigma \left(w^2 + 1\right) &= \text{constant}, \label{shock2}
\end{align}
which apply when the magnetic forces are small. These equations give the
non-trivial jump conditions: 
\begin{align}
    w_d &= \frac{1}{w_u} \label{jump1} \\
    \text{and } \sigma_d &= \sigma_uw_u^2, \label{jump2} 
\end{align}
where $u$ and $d$ denote the upstream and downstream sides of the shock; these
are also used for any subshocks downstream of the principal shocks. KK02 also
derived jump conditions for use when the magnetic forces are important; these
are not required in this work. It is possible that as the positive Hall
parameter becomes larger the magnetic diffusion subshock may require different
jump conditions that take the twisting of the magnetic field lines due to the
Hall effect into account.  

The magnetic field strength does not change in the centrifugal shock front as
the magnetic pressure and tension terms are not large enough to change the
field behaviour. In the post-shock region the field increases as the other
variables settle to the asymptotic behaviour in \S\ref{innerbc}. While the
magnetic field is unchanged by the passage of the shock, its position and
strength depend upon the Hall parameter. The increased magnetic diffusion can
also cause the formation of subshocks -- rings of sharply-enhanced density in
the post-shock region -- as the outward-moving flux causes the infalling gas
to be slowed. 

As with the magnetic diffusion subshock, the centrifugal subshocks are
preceded by a sonic point that must be calculated manually and the subshock
position is found using the binary search above. Inwards of the centrifugal
shocks the variables approach the asymptotic behaviour with some overshoots
and corrections. 

\subsection{Innermost integration}\label{num:inner}

In the innermost regions of the collapse the derivatives $d\sigma/dx$ and
$db_z/dx$ become very large in comparison to the other derivatives. These can
correspondingly cause small numerical errors in the calculation of these
derivatives and their integrals to build and trigger the appearance of a
spontaneous singularity where the variables diverge dramatically from the
asymptotic solution \citep[discussed in][]{l1998}. This behaviour is avoided
by replacing the full MHD equations with a simplified set that can be
integrated all the way to the inner boundary: the problematic derivatives are
replaced by approximations derived from the expected asymptotic behaviour,
that is,  
\begin{align}
    \frac{db_z}{dx} &= -\frac{5}{4}\frac{m_c^{3/4}}{\sqrt{2\delta}}\,x^{-9/4} 
     \label{simpdbz}\\
    \text{and }\frac{d\sigma}{dx} &= -\frac{3}{2}\,
     \frac{\sqrt{2m_c}f}{2\delta\sqrt{(2\delta/f)^2 + 1}}\,x^{-5/2}.
      \label{simpdsigma}
\end{align}
These are then substituted into the other equations so that $\sigma$, $b_z$,
$b_{\phi,s}$ and $h$ are found by solving Equations \ref{sscrm}, \ref{ssvhe},
\ref{ssin} and \ref{ssb_phis} simultaneously, while the remaining equations
are integrated to the boundary. 

Switching between the full and simplified models occurs after the minimum in
the surface density as the variables follow the asymptotic solution, but
before they diverge from the expected behaviour. Typically this occurs when
the old and new values of $\sigma$ match to within 0.01/$x$ ($\sim 0.1$ 
per cent of the original value) and $d\sigma/dx$ calculated using both methods
matches less well to around 200/$x$ ($\sim 7$ per cent). When the precision of
the match is as high as possible the change between the two equation sets is
smooth; the required precision of the match is raised when the switch is
apparent as this clearly influences the accuracy of the calculations.

The simplified set of equations is not subject to the same numerical
instabilities as the full set unless the guess at the matching point is
particularly poor. Should such a guess be adopted then the routine is unable
to match onto the simplified model, and the failure of the inwards integration
is used to refine the values at the matching point. 

The integration is continued to the inner boundary (typically at $x_\text{in}
= 10^{-4}$) where the variables are compared to the inner conditions given by
Equations \ref{in-m} and \ref{in-psi}. The differences between the expected
and integrated variables are passed back to the shooting routine, which
modifies the values at $x_m$ and begins the next iteration. The similarity
solution is considered to have converged when the integrated variables match
the boundary conditions to at least 0.002 per cent. Due to the ease of
convergence in the outwards direction, the outer variables match to a much
higher degree. 

\section{Results}\label{results}

Superficially, the similarity solutions with Hall diffusion are similar to the
fiducial ambipolar diffusion solution of KK02. We match their boundary
conditions and parameters, listed in Table \ref{tab-bc}, holding the ambipolar
diffusion parameter constant at $\tilde{\eta}_A = 1.0$ so that the only
changes between the similarity solutions presented in their work and ours are
those wrought by the addition of Hall diffusion. 
\begin{table}
\caption{Boundary conditions and parameters.}
\begin{center}
\begin{tabular}{lcr}
\hline
\multicolumn{2}{c}{parameter} & value\\
\midrule
 ambipolar diffusion & $\tilde{\eta}_A$ & 1.00 \\
 cap on $b_{\phi,s}$ & $\delta$		& 1.00 \\
 magnetic braking    & $\alpha$	  	& 0.08 \\
\hline
\multicolumn{2}{c}{boundary condition} & value \\
\midrule
 mass-to-flux ratio 	& $\mu_0$ & 2.90 \\
 rotational velocity 	& $v_0$	  & 0.73 \\
 radial velocity 	& $u_0$	  & $-1.00$ \\
 core accretion rate 	& $A$	  & 3.00 \\
 \bottomrule
 \end{tabular}
\end{center}
\label{tab-bc}
\end{table}

\subsection{Ambipolar diffusion solution}

\begin{figure}
  \centering
  \includegraphics[width=84mm]{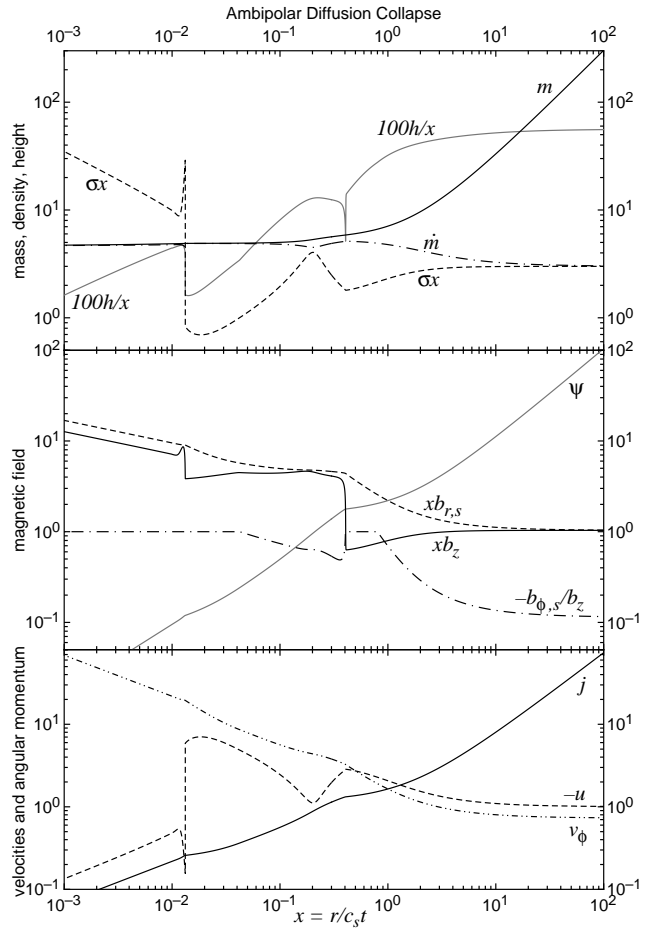}
  \caption{Calculated similarity solution for collapse with only ambipolar
diffusion, duplicating KK02's fig.\ 7. The enclosed mass, accretion rate,
surface density and scale height are displayed in the upper panel; the 
magnetic field components and enclosed flux in the centre panel; and the gas
velocities and angular momentum in the lower panel. The parameters and
boundary conditions are as in Table \ref{tab-bc}; the central mass is $m_c = 
4.67$; and the magnetic diffusion and centrifugal shocks are located at $x_d =
0.406$ and $x_c = 1.32 \times 10^{-2}$ respectively.}
\label{ad}
\end{figure}

As a test of the computational code we calculate the fiducial ambipolar
diffusion solution of KK02 in Fig.\ \ref{ad}, and follow their discussion of
the solution here. The outer regions of all the solutions match IMHD
similarity solutions, as the mass-to-flux and mass-to-angular momentum 
ratios are constant while the material falls in at supersonic speeds. The
radial velocity and scale height are dominated by the self-gravity of the
disc, which causes the material to fall towards the midplane before it is
accelerated towards the central mass. The magnetic field gradually builds up
as the matter falls inward, becoming important to the dynamics at around $x
\approx 2$ where magnetic braking starts to affect the angular momentum
transport and the constant ratio of the mass-to-angular momentum breaks down.
The azimuthal field attains its capped value, and becomes important to the
angular momentum transport. 

The mass and angular momentum in this region tend towards plateau values: the
mass to that in Equation \ref{mcguess}, and the angular momentum to 
\begin{equation}
    j_{pl} \approx \frac{m_{pl}v_0}{A}, \label{jguess}
\end{equation}
which depends upon the boundary conditions. This plateau forms as the gravity
of the central point mass starts to dominate and the gas starts to fall inward
rapidly. The surface density and magnetic field build up and ambipolar
diffusion becomes important, causing flux-freezing to break down at the
magnetic diffusion shock (at $x_d = 0.406$).

\begin{figure}
  \centering
  \includegraphics[width=80mm]{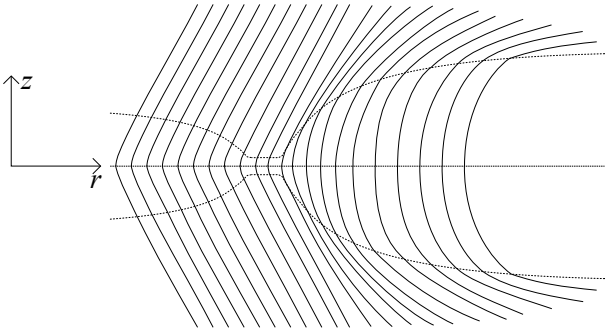}
  \caption{Schematic of the poloidal magnetic field in the magnetic diffusion
shock. The disc (dotted lines) is compressed as the vertical field becomes
large, causing the field lines at the surface to straighten from being largely
radial upstream to having roughly equal poloidal components downstream. (Not
to scale.)} 
\label{fig-adsilverfish}
\end{figure}

This shock takes the form of a sudden increase in the vertical magnetic field 
as the field lines diffusing against the flow in the downstream ambipolar
diffusion-dominated regime meet those coming inward under IMHD collapse. The
field geometry is illustrated in Fig.\ \ref{fig-adsilverfish}: upstream the
surface magnetic field is dominated by the radial component, which is an order
of magnitude larger than the vertical and azimuthal components. During the
shock (and the downstream transition region) the field lines straighten until
the poloidal components are approximately equal at the disc surface. The
increase in field strength slows the gas, and magnetic squeezing comes to 
dominate the vertical compression. 

Immediately interior to the magnetic diffusion shock the poloidal field
components scale with $\sim x^{-1}$ as the surface density and thickness of
the disc increase and the gas is slowed by the radial magnetic pressure.
Rotation is not dynamically important and the shock has a similar structure to
that in the nonrotating similarity solutions of \citet{cck1998}. The gravity
of the central mass becomes the dominant radial force on the gas at the end of
the post-shock region (the peak in the surface density) and the matter is
accelerated inwards until it is in near-free fall collapse. 

As in slowly-rotating similarity solutions that are non-magnetic
\citep[e.g.][]{sh1998} or include IMHD (e.g.\ KK02 \S3.2) rotation remains
dynamically unimportant until the vicinity of the centrifugal shock. Between
the two shocks the gravity of the central mass dominates the radial infall of
the gas, which is slowed only a little by ambipolar diffusion of the magnetic
flux and the magnetic pressure. The magnetic braking increases the azimuthal
field component until it is again capped; the activation of the cap causes the
change in the behaviour of the scale height at $x \sim 4.5 \times 10^{-2}$ as
$b_{\phi,s}$ then contributes more to the vertical squeezing forces. The
enclosed mass and accretion rate flatten and remain near-constant throughout
the remainder of the collapse. The angular momentum starts to plateau again
and the centrifugal force becomes large and equal to the gravitational force,
triggering the centrifugal shock at $x_c = 1.32 \times 10^{-2}$. 

The centrifugal shock slows the gas so that the infall is subsonic and the
surface density increases by more than an order of magnitude. The shock is
followed by a thin layer in which the azimuthal and vertical magnetic field
components increase rapidly. This causes the angular momentum to drop to its
asymptotic behaviour, as the surface density and the other variables adjust
with a few overshoots towards their expected rotationally-supported disc
behaviour. The transition between the full model and the simplified set of
equations occurs at $x \sim 8.6 \times 10^{-3}$, after the variables have
joined onto the asymptotic disc described by Equations
\ref{in-m}--\ref{in-bphis}. 

The Keplerian disc itself is small, with a mass $\sim 5$ per cent that of the
central point mass. The nondimensional mass at the origin is $m_c = 4.67$,
corresponding to a moderate accretion rate of $\dot{M}_c = 7.6 \times 10^{-6}$
M$_\odot$ yr$^{-1}$ (so that at a time $t = 10^5$ yr, the central mass is $M_c
= 0.76$ M$_\odot$). The surface density of the disc depends upon the ambipolar
diffusion and azimuthal field cap parameters, as does the infall velocity,
which is subsonic and very low. The disc is extremely thin, and the vertical
squeezing is dominated by the tidal and self-gravitational forces. The
magnetic flux in the disc scales as $x^{3/4}$, so that $\psi \to 0$ as $x
\to 0$; clearly the amount of flux present in the protostar depends upon more
detailed flux transport and destruction mechanisms than are included in this
model, such as Ohmic diffusion \citep[e.g.][]{lm1996} and reconnection
\citep[e.g.][]{gs1993b, l2005}.

\subsection{Hall diffusion solutions}

\begin{figure}
  \centering
  \includegraphics[width=84mm]{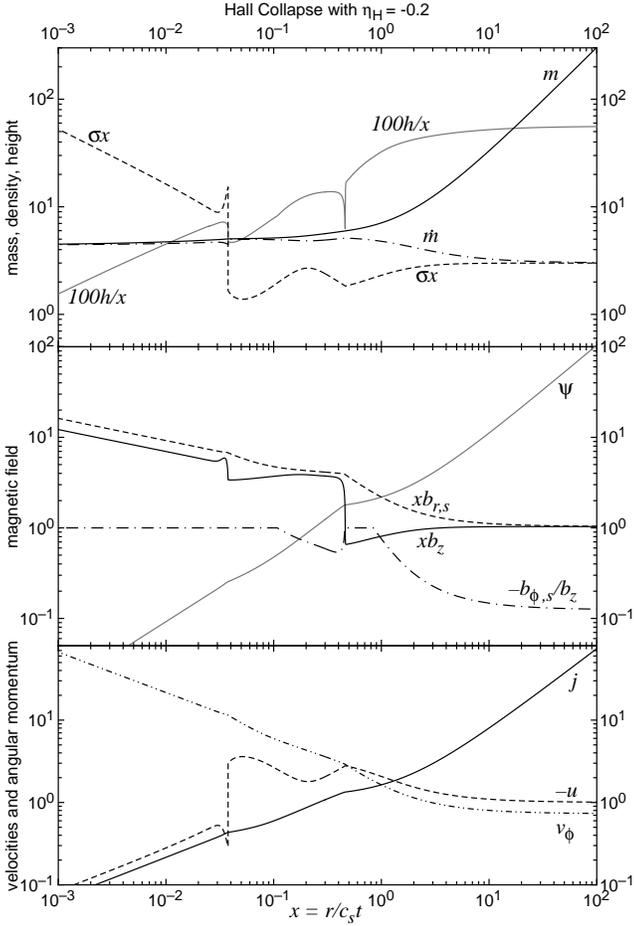}
  \caption{Hall collapse with $\tilde{\eta}_H = -0.2$. The variables are as in
Fig.\ \ref{ad}; the parameters and boundary conditions are given in Table
\ref{tab-bc}. The central mass is $m_c = 4.23$; the magnetic diffusion and
centrifugal shocks are located at $x_d = 0.461$ and $x_c = 3.78 \times
10^{-2}$ respectively; these are increased from the non-Hall positions, and
the post-shock regions are smoothed by Hall diffusion.} 
\label{fig-hall-0.2}
\end{figure}

The first of the similarity solutions with Hall diffusion is that presented in
Fig.\ \ref{fig-hall-0.2} for the self-similar collapse of a molecular cloud
with $\tilde{\eta}_H = -0.2$ and other parameters matching those in Fig.\
\ref{ad}. The negative Hall parameter solutions have more radial diffusion of
the magnetic field against the neutral fluid and charged grains, so that the
magnetic pressure builds up earlier in the collapse process, triggering
the formation of the magnetic diffusion shock. The negative Hall parameter
also increases the initial rate of magnetic braking so that $b_{\phi,s}$
attains its capped value earlier in the collapse, and the magnetic
braking is then determined by the strength of the vertical field component. 

As in the no-Hall solution, at the outer edge of the collapse in Fig.\
\ref{fig-hall-0.2} the matter is falling in supersonically under IMHD. As the
surface density builds up the field does too, causing the magnetic pressure
and tension terms to become important, while the magnetic braking transports
angular momentum from the infalling gas to the external envelope. The angular
momentum and enclosed mass start to plateau as the dominant force on the
radial velocity switches from the self-gravity of the disc to the gravity of
the central mass, which in turn causes the accretion rate to taper off. The
formation of the magnetic diffusion shock at $x_d = 0.461$ (increased from the
non-Hall solution) is caused by the decoupling of the field from the neutral
particles. 
 
The magnetic diffusion shock in this solution is weaker than in Fig.\ \ref{ad}
as most of the neutral particles and grains have already decoupled from the
magnetic field, so that the vertical field component increases by only 4.5
times (\textit{cf.} 6.2 times in Fig.\ \ref{ad}). The disc is less vertically
compressed by the field, producing a thinner shock, and $b_{\phi,s}$ does not
grow as rapidly. Within the shock the field is further decoupled from the
neutrals and grains, allowing Hall and ambipolar diffusion to become more
important downstream of the shock and throughout the remainder of the
solution. The field lines straighten as in Fig.\ \ref{fig-adsilverfish};
although the radial field component is still dominant, the vertical component
increases in the shock until it is just smaller than $b_{r,s}$. 

Downstream of the magnetic diffusion shock the surface density gradually
increases as the infall velocity is slowed by the larger magnetic support.
This post-shock region is smoother than that without Hall diffusion, 
presenting a gentler transition to the free fall collapse that occurs outside
of the rotationally-supported disc. The vertical field scales as $x^{-1}$ in
this region as the increased radial diffusion means that there are fewer field
lines in total moving against the flow of the neutral particles. The magnetic
braking decreases the angular momentum efficiently until $b_{\phi,s}$ attains
its capped value and $j$ begins to plateau once more. 

The centrifugal force builds up and triggers the centrifugal shock at $x_c =
3.78 \times 10^{-2}$ (\textit{cf.}\ $x_c = 1.32 \times 10^{-2}$ in Fig.\
\ref{ad}). The shock is a discontinuity in the surface density and radial
velocity, which is again less strong than in the solution without Hall
diffusion, and inwards of this the vertical and azimuthal field components
increase steeply as the field reacts to the shock (although $b_{\phi,s}$
remains capped at $-\delta b_z$). Downstream the variables tend (with
overshoots) towards their asymptotic values. 

The inner disc is in Keplerian rotation satisfying Equations
\ref{in-m}--\ref{in-f}. The central mass is $m_c = 4.42$, decreased from the
non-Hall solution, and corresponds to an accretion rate of $\dot{M}_c = 7.21
\times 10^{-6}$ M$_\odot$ yr$^{-1}$. The scalings of the other variables with
respect to $x$ are the same as in Fig.\ \ref{ad}, however the surface density
is increased by the larger magnetic diffusion parameter $f = 1.72$
(\textit{cf.} $f(\tilde{\eta}_H = 0) = 1.\dot{3}$); and the increased radial
magnetic diffusion causes the strength of the magnetic field to be decreased
from that in the ambipolar diffusion-only solution. This in turn means that
less matter can lose its angular momentum and fall onto the central mass, so
that the gas is at a higher surface density in this larger Keplerian disc. 

\begin{figure}
  \includegraphics[width=84mm]{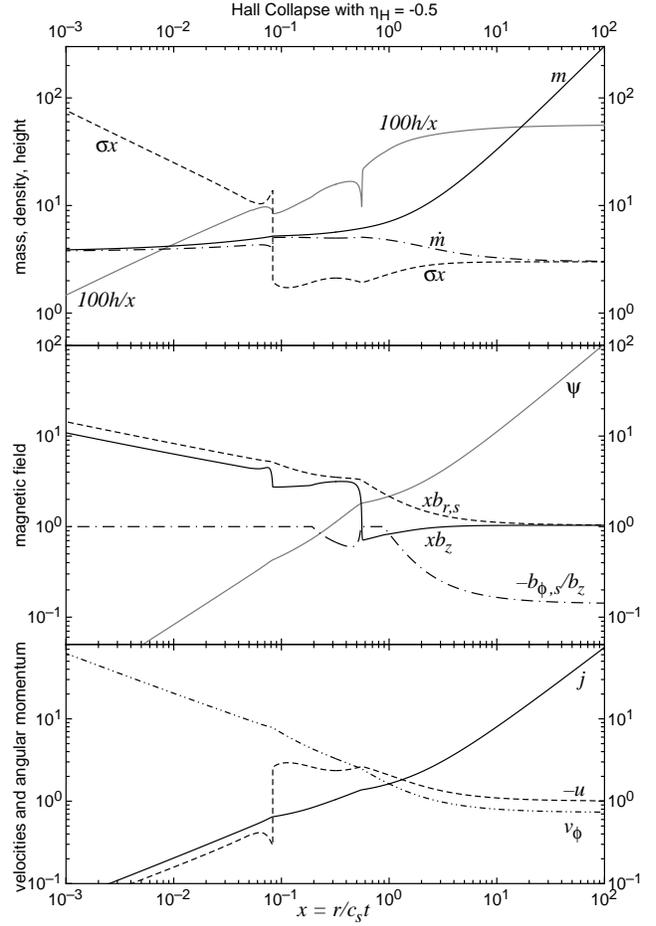}
  \caption{Collapse with larger Hall parameter $\tilde{\eta}_H = -0.5$. The 
boundary conditions and parameters match those in Fig.\ \ref{fig-hall-0.2}
(Table \ref{tab-bc}). The central mass is reduced to $m_c = 3.77$ and the
magnetic diffusion and centrifugal shocks moved outwards to $x_d = 0.557$ and
$x_c = 8.31 \times 10^{-2}$ as the increased magnetic diffusion smooths the
post-shock regions and increases the size of the Keplerian disc.} 
\label{fig-hall-0.5}
\end{figure}
The next similarity solution, presented in Fig.\ \ref{fig-hall-0.5}, shows the
calculation with $\tilde{\eta}_H = -0.5$ on the same scale and with the same
parameters as Fig.\ \ref{fig-hall-0.2}. The total radial magnetic diffusion is
further increased so that many of the neutral particles have decoupled from
the field before the magnetic diffusion shock at $x_d = 0.557$; this causes
the intensity of the shock to drop further so that the vertical field strength
is only 4 times larger than its original value. There is less of a magnetic
wall at this point as less flux remains to be decoupled from the neutrals
within the shock itself. 

As in the previous solution with negative Hall parameter the post-magnetic
diffusion shock region is smoothed, with even less change in the surface
density and radial velocity. The gas is slowed by the magnetic diffusion
shock, but the gravity of the central mass quickly overcomes this and pulls
the fluid inwards. The radial velocity downstream of the post-shock region
increases as the fluid nears the protostar, however it remains below the free
fall velocity at all times. The mass and angular momentum both plateau in this
region before the increasing centrifugal force triggers the centrifugal shock. 

The centrifugal shock occurs much earlier in this similarity solution at $x_c
= 8.31 \times 10^{-2}$. This is brought about by the decreased values of $b_z$
and $b_{\phi,s}$ in the free fall region, which reduce the amount of magnetic
braking that takes place and cause the centrifugal force to become important
earlier. Inwards of this shock is a much wider region of adjustment as the
variables join the inner disc solution. 

\begin{figure}
  \includegraphics[width=84mm]{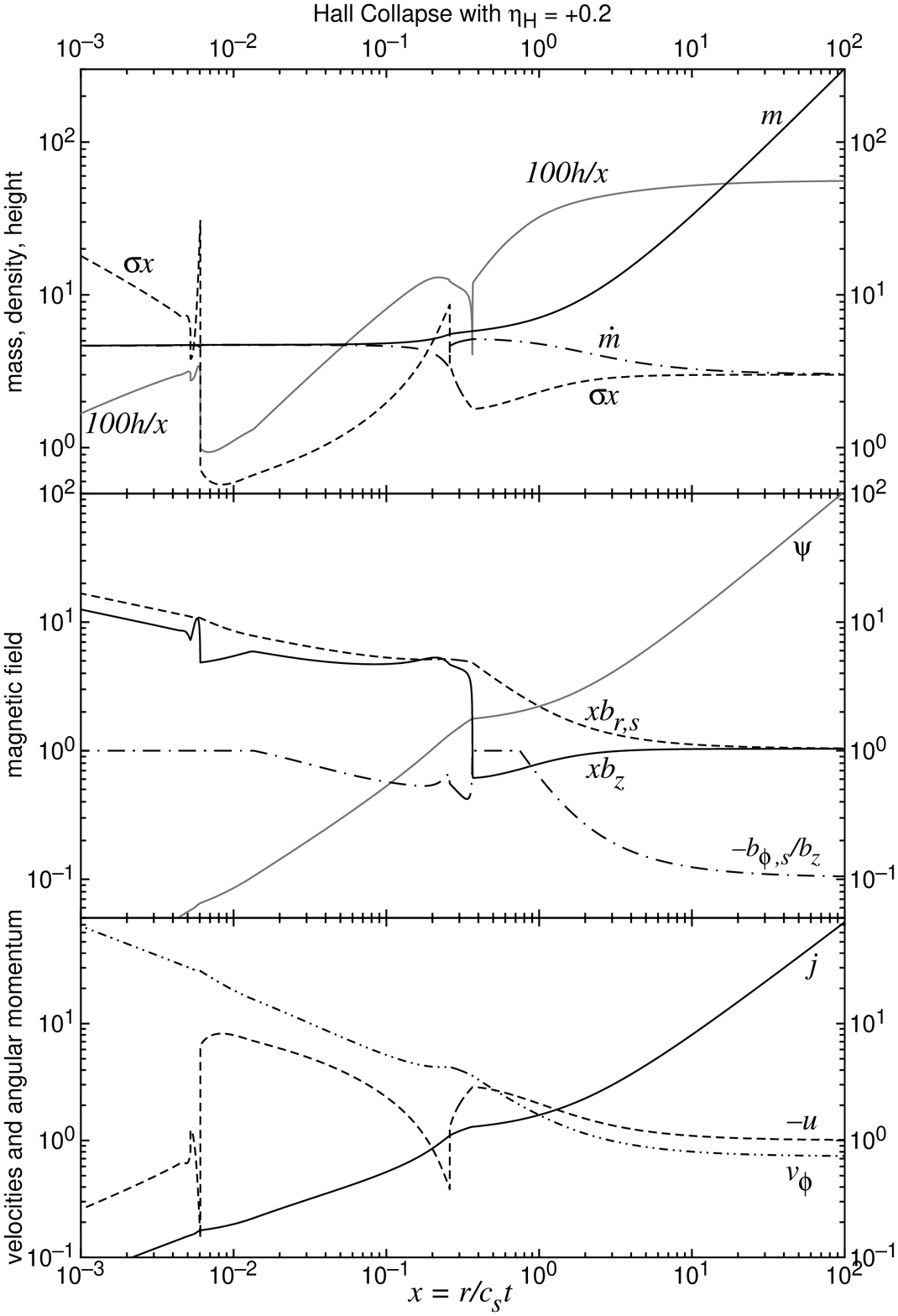}
  \caption{Gravitational collapse with positive Hall parameter $\tilde{\eta}_H
= +0.2$. The boundary conditions and parameters match Fig.\ \ref{fig-hall-0.2}
and Table \ref{tab-bc}; the central mass is $m_c = 4.63$. The positive Hall
term causes the formation of subshocks: the magnetic diffusion shocks are
located at $x_d = 0.365$ and $x_{d2} = 0.260$; the centrifugal shocks at $x_c
= 6.05 \times 10^{-3}$ and $x_{c2} = 5.21 \times 10^{-3}$.} \label{fig-hall0.2}
\end{figure}
The Keplerian disc is substantially larger than that in the previous solution,
containing $\sim 38$ per cent of the mass of the central protostar. The
surface density has also increased as the magnetic diffusion parameter is $f =
2.31$, while the lowered central mass $m_c = 3.77$ corresponds to a central
accretion rate of $\dot{M}_c = 6.15 \times 10^{-6}$ M$_\odot$ yr$^{-1}$.
Again, the larger disc corresponds to a lower accretion rate, as the reduced
magnetic braking prevents the fluid from losing rotational support and falling
in.  

The final solution presented here is that with $\tilde{\eta}_H = +0.2$ in
Fig.\ \ref{fig-hall0.2}, which is the most dynamically different from those of
KK02. Although the initial conditions and parameters match those in Fig.\
\ref{fig-hall-0.2}, the change in the sign of the Hall parameter, which
corresponds to a reversal of the orientation of the magnetic field with
respect to the direction of rotation, introduces many changes to the collapse
dynamics. 

These begin at the magnetic diffusion shock, which has moved inwards from
$\tilde{\eta}_H = 0$ solution to $x_d = 0.366$. This shock is of increased
intensity due to the reduced radial magnetic diffusion upstream, which causes
a larger increase in $b_z$ in the shock. The magnetic braking downstream is
increased by the presence of a stronger field and the sign of the Hall term in
Equation \ref{ssb_phis}. The disc is more sharply compressed as the field
lines straighten at the shock front, and the fluid is so slowed by the
increase in the magnetic pressure that a second shock front forms at $x_{d2} =
0.260$. 

\begin{figure}
  \centering
  \includegraphics[width=69mm]{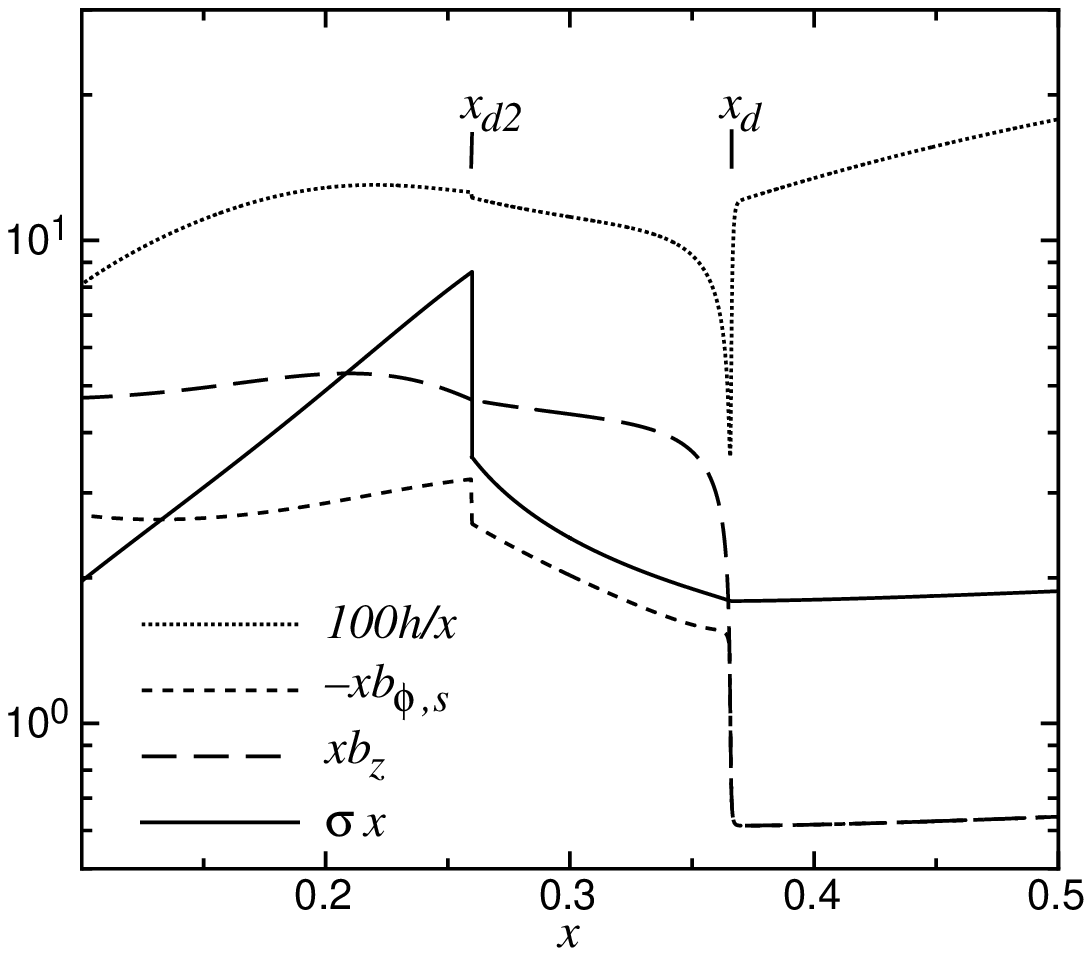}
  \caption{The magnetic diffusion shock and subshock $x_{d2}$, for the
collapse with $\tilde{\eta}_H = +0.2$. As seen in Fig.\ \ref{fig-hall0.2}, $j$
and $v_\phi$ are continuous, and $-u$ mirrors the behaviour of $\sigma x$.} 
\label{fig-xdshocks}
\end{figure}
In the magnetic diffusion subshock, shown at higher resolution in Fig.\
\ref{fig-xdshocks}, the fluid is slowed until the radial velocity is low and
subsonic. The surface density increases under the jump conditions from \S
\ref{num:jump}; this ring of matter contains approximately 18 per cent of the
protostellar mass. The azimuthal field component and the scale height also
increase, while $db_z/dx$ decreases steeply. The infall region downstream of
the subshock is wider in logarithmic similarity space than in the previous
solutions, with the increased magnetic braking reducing the angular momentum
more quickly as the radial velocity increases. The surface density drops as
the fluid falls in and magnetic squeezing dominates the vertical compression
until the gravity of the central mass takes over near the centrifugal shock. 

\begin{figure}
  \centering
  \includegraphics[width=69mm]{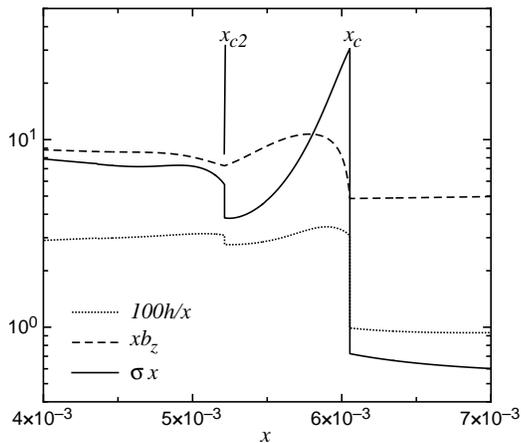}
  \caption{The centrifugal shock and subshock (denoted $x_{c2}$), for the
$\tilde{\eta}_H = +0.2$ solution on a linear $x$ scale. Again, $j$ and
$v_\phi$ are continuous, while $-u$ mirrors $\sigma x$ as in Fig.\
\ref{fig-hall0.2}.}  
\label{fig-xcshocks} 
\end{figure}
The centrifugal shock occurs at $x_c = 6.05 \times 10^{-3}$, half that of the
$\tilde{\eta}_H = 0$ solution, dramatically decreasing the size of the
rotationally-supported disc. This change is brought about by the increased
magnetic braking caused by the positive Hall parameter, which reduces the
angular momentum so that the centrifugal force is not dynamically important
until the gas is very close to the protostar. Downstream of the shock the
slowed fluid accelerates inwards as the magnetic field increase forces
additional magnetic braking and a drop in the centrifugal force. The surface
density drops as the infall velocity becomes supersonic, and the centrifugal
force becomes important once more, triggering the subshock at $x_{c2} = 5.21
\times 10^{-3}$ which is shown in Fig.\ \ref{fig-xcshocks}. The matter is
slowed in the subshock until it is again subsonic. Although the surface
density increases in the subshock, the disc does not become
gravitationally-unstable as the Toomre $Q$ parameter \citep{t1964} remains
above $8$ in this region.

Downstream of the subshock the variables again settle with overshoots to the
Keplerian disc behaviour. The central mass is $m_c = 4.63$, corresponding to a
protostellar accretion rate of $\dot{M}_c = 7.53 \times 10^{-6}$ M$_\odot$
yr$^{-1}$, while the disc contains only $1.6$ per cent the protostellar mass.
The magnetic diffusion parameter is $f = 0.945$, reflecting the decrease in
surface density as the magnetic field strength increased.

The magnetic diffusion and centrifugal subshocks both occur only in the
calculations where $\tilde{\eta}_H$ is positive, as the larger magnetic
pressures and braking caused by the increase in the infalling magnetic field
force the gas to rapidly change in radial velocity and density. The number of
centrifugal subshocks increases with $\tilde{\eta}_H$ --- three subshocks have
been observed in one similarity solution that was not properly converged (due
to the numerical instability of the subshocks) at the time of publication ---
while only one magnetic diffusion subshock has been observed. 

The diffusion parameter of the disc, $f$, cannot be less than or equal to zero
in our collapse solutions, which do not contain counter-rotation. This
restriction limits the range of positive $\tilde{\eta}_H$ that can be
explored: as $\tilde{\eta}_H$ increases, the size and surface density of the
rotationally-supported disc decrease, and the rings of gas formed by the
subshocks are more likely to be gravitationally unstable. Similar constraints
were found to limit the launching of disc winds in the analysis of
\citet{skw2011}, who showed that disc wind solutions only exist for particular
combinations of the field polarity and the ratio of the Hall to ambipolar
diffusion parameters. 

\begin{figure}
  \centering
  \includegraphics[width=83mm]{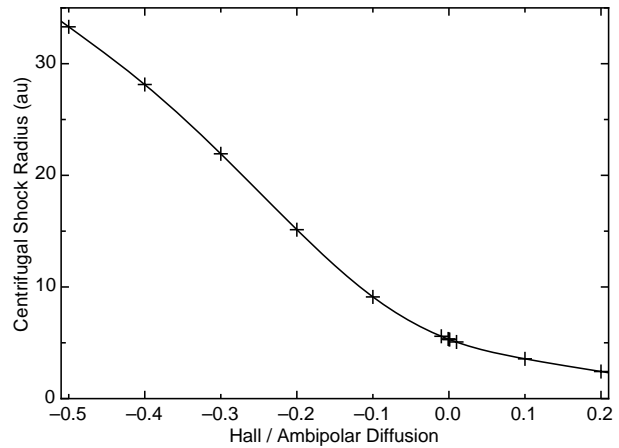}
  \caption{The centrifugal shock radius (in au at $10^4$ years) against the
ratio of the Hall to ambipolar diffusion parameters.} 
\label{fig-rc}
\end{figure}
The positions of the shocks change with the Hall parameter as demonstrated in
Fig.\ \ref{fig-rc}, which plots the dimensional centrifugal shock position for
similarity solutions with parameters as in Table \ref{tab-bc} at $10^4$ years
\citep[illustrated in][]{b2011} against the ratio of the Hall to ambipolar
diffusion parameters. While both directions of Hall drift contribute to the
size of the disc, the radial drift of $B_z$ (which increases when
$\tilde{\eta}_H < 0$) has a greater effect on the radius of the centrifugal
shock than the azimuthal Hall drift. 

\begin{figure}
  \centering
  \includegraphics[width=83mm]{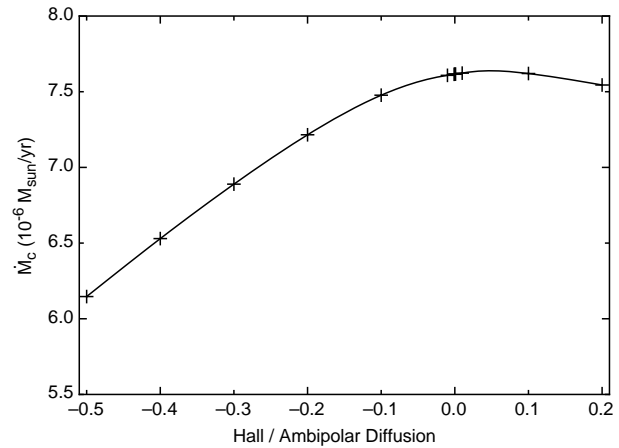}
  \caption{The protostellar accretion rate (in $10^{-6}$ M$_\odot$/yr) against
the ratio of the Hall to ambipolar diffusion parameters.} 
\label{fig-mdot}
\end{figure}
There also exists a correlation between the accretion rate onto the central
protostar and the radial magnetic field diffusion, shown in Fig.\
\ref{fig-mdot} \citep[see also][]{cck1998}. The accretion depends upon the
disc radius, with larger discs corresponding to lower accretion rates and vice
versa \citep{asl2003b}, as a more negative Hall parameter causes increased
drag on the neutrals and reduces the radial velocity of the fluid. The disc
radius also depends on the initial rotational velocity, as the centrifugal
force is more important and a larger disc forms when the initial angular
momentum of the core is large (KK02). 

The protostellar accretion rate appears to turn over at around $\dot{M}_c =
7.6 \times 10^{-6}$ M$_\odot$ yr$^{-1}$ as the ratio of the Hall to ambipolar
diffusion parameters becomes positive and greater than 0.05. This is due to
the formation of subshocks in the solutions with $\tilde{\eta}_H > 0$: as the
density is enhanced accretion through the disc drops. As the Hall parameter
increases further subshocks are introduced, the diffusion parameter tends
towards zero and the rings formed by the subshocks become more unstable. 

\section{Discussion}\label{discuss}

The similarity solutions clearly show that Hall diffusion changes the
structure and dynamics of the collapse of molecular cloud cores into
protostars and protostellar discs. The rotationally-supported disc size, and
the accretion rate onto the protostar are determined by the ratio of the Hall
and ambipolar diffusivities, which influences the magnetic braking affecting
the rotation of the collapsing core. It is also clear that Hall diffusion can
inhibit disc formation by enhancing the magnetic braking, or by counteracting
ambipolar diffusion to the point that the field starts to infall faster than
the fluid, increasing the magnetic pressure and tension.
\begin{table*}
\centering
\begin{minipage}{145mm}
\caption{The surface density and vertical field component in the Keplerian
disc at $r = 1$ au, the protostellar mass and the size and mass of the
Keplerian disc at $t = 10^4$ years for the solutions depicted in Fig.\
\ref{fig-comp}.}\label{tab-Sigma} 
\centering
\begin{tabular}{rrrllr}
\toprule
 $\tilde{\eta}_H$ & $\Sigma$ (g cm$^{-2}$) & $B_z$ (G) &
	$M_c$ (M$_\odot$) & $M_{disc}$ (M$_\odot$) & $R_c$ (au)\\
 \midrule
 $-0.2$	    & $1\,920\quad$ & $0.289\;\;$ & $7.21 \times 10^{-2}$ & $9.99 \times
	10^{-3}$ & $15.10\quad$ \\ 
 $ 0\;\;\;$ & $1\,250\quad$ & $0.299\;\;$ & $7.62 \times 10^{-2}$ & $3.75 \times
	10^{-3}$ & $5.31\quad$ \\ 
 $ 0.2$     & $620\quad$   & $0.304\;\;$ & $7.54 \times 10^{-2}$ & $1.24 \times
	10^{-3}$ & $2.43\quad$ \\ 
 \bottomrule
 \end{tabular}
\end{minipage}
\end{table*}

\begin{figure}
  \centering
  \includegraphics[width=84mm]{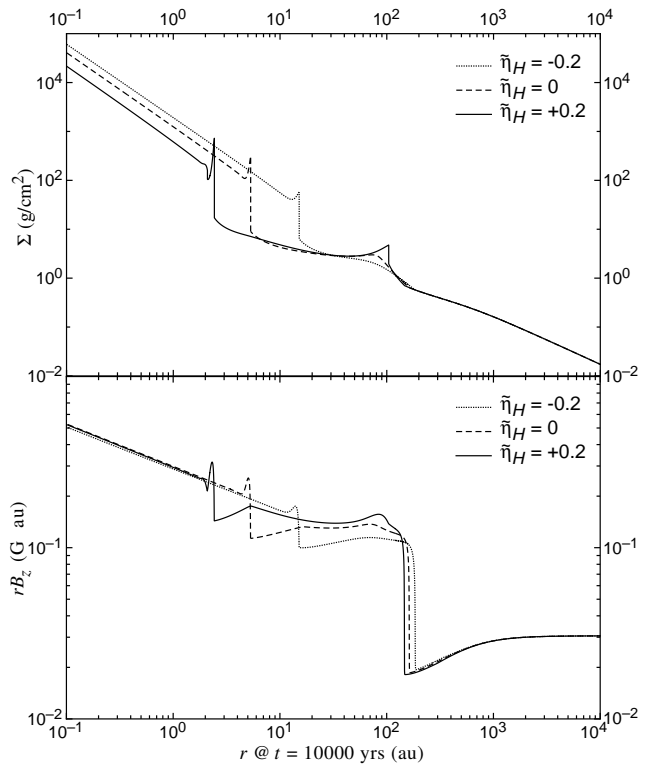}
  \caption{The surface density $\Sigma$ and the vertical magnetic field
component $B_z$ plotted against radius at $t = 10^4$ yr for the solutions with
$\tilde{\eta}_A = 1.0$ and $\tilde{\eta}_H = -0.2$ (dotted line), $0$ (dashed
line) and $+0.2$ (solid line). These solutions were plotted individually in
nondimensional form in Figs.\ \ref{fig-hall-0.2}, \ref{ad} and
\ref{fig-hall0.2} respectively.} 
\label{fig-comp}
\end{figure}

The dependence of the similarity solutions on the orientation of the magnetic
field and the sign of the Hall diffusion parameter $\tilde{\eta}_H$ (more
specifically upon the sign of $\tilde{\eta}_H (\mathbf{B}\cdot\Omega)$) gives
rise to two different patterns of collapse behaviours. The similarity
solutions with $\tilde{\eta}_H = 0$ and $\pm 0.2$ are converted to dimensional
form and plotted against the radius $r$ (at $t = 10^4$ years) in Fig.\
\ref{fig-comp}, with the surface density in the upper panel and the vertical 
magnetic field strength plotted as $rB_z$ in the lower. The solutions have the
same boundary conditions and parameters (Table \ref{tab-bc}), and the surface
density and vertical field strength (at $r = 1$ au), the central mass, and the
mass and size of the inner disc (all at $t = 10^4$ years) are listed in Table
\ref{tab-Sigma}. The outer regions where IMHD holds are near-identical and it
is only near the magnetic diffusion shock at $r \approx 100$ au that the
changes brought on by Hall diffusion become apparent. 

The dotted lines in Fig.\ \ref{fig-comp}, corresponding to the negative Hall
solution in Fig.\ \ref{fig-hall-0.2}, show the formation of a large
rotationally-supported disc that has radius $R_c \approx 15$ au at ($t = 10^4$
years) and the highest inner Keplerian disc surface density of all the
solutions. The dashed lines are the $\tilde{\eta}_H = 0$ solution from Fig.\
\ref{ad}, which possesses a disc radius half that the negative Hall solution.
The surface density of the Keplerian disc has decreased by a constant factor
from that in the negative Hall solution. Finally, the solid curves
characterise the similarity solution with positive Hall parameter
$\tilde{\eta}_H = 0.2$ (Fig.\ \ref{fig-hall0.2}) which has a Keplerian disc
that is almost an order of magnitude smaller than that in the negative Hall
case. This disc is bounded by a thin ring of enhanced density that rapidly
drops off as the magnetic field peaks; the material is then shocked again and
comes to match onto the inner solution. The density is much lower than in
previous solutions, and the disc grows at a slower rate. 

The similarity solutions span many orders of magnitude in both radius and
density, and the inclusion of a Hall parameter that is 20 per cent that of the
ambipolar diffusion parameter has a large effect on the behaviour of the
magnetic field. In the intermediate region between the magnetic diffusion
shock (at $R_d \approx 100$ au in Fig.\ \ref{fig-comp}) and the centrifugal 
shock ($R_c \approx 1-10$ au), the azimuthal field tension causes the Hall
drift to enhance the radial diffusion of the field lines when $\tilde{\eta}_H$
is negative. The magnetic diffusion shock occurs earlier in the collapse and
is less dynamic than in the other solutions, as much of the field has already
been decoupled from the fluid. 

However, when $\tilde{\eta}_H$ is positive then Hall diffusion acts to reduce
the net radial diffusion, resulting in magnetic walls and subshocks that
disrupt the flow. The magnetic field carried inwards increases, and the
magnetic pressure and tension terms remain important throughout the collapse.
Any twist in the field lines causes an increase in the magnetic pressure
gradient, so that the net amount of radial diffusion drops off as the magnetic
braking slows the rotation. 

There is a similar duality to the azimuthal field drift. Again looking at the
region between the two shocks, when $\tilde{\eta}_H$ is negative Hall drift
occurs in the azimuthal direction, twisting up the field lines in the
pseudodisc and creating a leading torque on the neutral rotation. The reduced 
value of $B_z$ causes the azimuthal field component to reach its capped value
$B_{\phi,s} = -\delta B_z$ sooner, and the magnetic braking, which depends
upon $B_zB_{\phi,s}$, is also reduced. Because of this less angular momentum
is removed from the pseudodisc, causing the centrifugal force to become
dynamically important earlier and a larger rotationally-supported disc to
form. 

In the other orientation when $\tilde{\eta}_H$ is positive, Hall and ambipolar
diffusion act together to untwist the field lines in the pseudodisc. In these
similarity solutions $B_z$ is larger, and so while it takes longer for
$B_{\phi,s}$ to achieve its capped value there is more magnetic braking and
the angular momentum is further reduced. A smaller Keplerian disc forms due to
the reduced centrifugal force, and both shocks have subshocks where the
magnetic forces alter the radial velocity of the fluid. Downstream of the
magnetic diffusion shock the radial magnetic pressure gradient slows the fluid
in the radial and azimuthal directions, while downstream of the centrifugal
shock the gas is accelerated inward as the increase in $B_z$ causes a burst of
magnetic braking that disrupts the disc and causes the formation of a
subshock. 

While the angular momentum behaviour between the shocks is changed by the
inclusion of Hall diffusion, the cap on $B_{\phi,s}$ acts to ensure that the
angular momentum in the inner disc is that expected for a Keplerian disc. The
cap, while physically motivated, replaces unspecified disc physics such as
reconnection, a disc wind, or turbulence, which would act to prevent the
azimuthal field component from greatly exceeding the vertical component; and
the magnitude at which it ought to act to limit $B_{\phi,s}$ is uncertain. It
is also unclear if such limiting of the azimuthal field component happens in
real collapsing cores, as numerical simulations do demonstrate tightly-wound
magnetic fields \citep[e.g.][]{mim2008}. The azimuthal field cap limits the 
similarity solution set explored to those in which discs form, however despite
this Hall diffusion has been shown to restrict disc formation in collapse
without counter-rotation if the Hall diffusion is too strong in comparison to
ambipolar diffusion and $\tilde{\eta}_H$ has the ``wrong'' sign. If the core
was initially rotating in the opposite direction, the opposite sign of
$\tilde{\eta}_H$ would be problematic to disc formation.

All of the solutions in Fig.\ \ref{fig-comp} and \citet{b2011} form protostars
of around $0.7$ solar masses with protostellar discs of radius $R_c
\sim$10--$150$ au and mass $M_d \sim 10^{-2}$--$10^{-1}$ M$_\odot$ in $t =
10^5$ years; these are the same order of magnitude expected from observations
of Class I YSOs \citep[e.g.][]{jetal2007}. The surface density of the disc is
quite sensitive to the Hall diffusion parameter, and scales as $\Sigma \propto
r^{-3/2}$ and $\Sigma \propto t^{1/2}$ in the inner Keplerian disc, with
values $\Sigma(r = 1$ au$, t = 10^4$ years$) \sim 10^3$ g cm$^{-2}$ (see Table
\ref{tab-Sigma}). These values of the surface density are consistent with what
is thought to have occurred in the solar nebula \citep[e.g.\ the minimum mass
solar nebula has $\Sigma = 1700$ g cm$^{-2}$ at $r = 1$ au;][]{w1977}.  

It has recently been argued that there exists a handedness to observations of
transverse gradients in the Faraday rotation measure across the base of jets
associated with active galactic nuclei \citep[AGN;][]{cckg2009}. The majority
of sources in which it was possible to determine the transverse gradients were
found to have clockwise gradients, implying that the outflow has a helical
magnetic field with a preferred magnetic polarity. One explanation of this
behaviour is that the Hall effect is important in the inner accretion disc,
acting to form a jet when the field has a positive magnetic polarity, and to
suppress jet formation when the polarity is negative \citep{k2010}. This
explanation fits the limited available data well, although it must be
confirmed by future observations at higher resolutions and sensitivities. 

Similarly, it may be possible to show the importance of the sign of the Hall
parameter observationally by measuring the polarisation of the magnetic field
with respect to the axis of rotation in Zeeman observations of newly-forming
stars and their discs. Should larger discs and lower accretion rates be
correlated with a particular field orientation then the Hall effect will have
been shown to affect the collapse process. ALMA (among other next-generation
instruments) shall be capable of imaging nearby dense prestellar cores and
their envelopes in both dust and molecular line emission, and could also be
used to observe polarised dust emission and map the magnetic field in cores.
Such observations could be sensitive enough to observe if there is any
difference in the field alignment between protostellar discs and their 
envelopes, and whether there is any correlation between disc size and the
direction of rotation in the disc. 

None of the solutions calculated the effects of very weak magnetic diffusion
or strong magnetic braking on the core. In numerical simulations with such
conditions the magnetic braking removes all of the angular momentum from the
fluid, preventing the formation of a rotationally-supported disc -- this
behaviour has been dubbed the ``magnetic braking catastrophe'' \citep[see 
e.g.][]{asl2003b, ml2008, ml2009, hc2009, lks2011}. Disc formation was assured
in this work by the cap placed upon $B_{\phi,s}$, which limits the twisting of
the field lines, however KK02 were able to demonstrate the magnetic braking
catastrophe by adopting a large $\delta$. Hall diffusion is capable of
inducing spin in an initially-nonrotating fluid \citep{wn1999}, and could
resolve the magnetic braking catastrophe in solutions with strong braking by 
spinning up the core in the opposite direction once magnetic braking has
removed the initial angular momentum. Hall-induced spin-up of an
initially-nonrotating core was demonstrated by \citet{kls2010-2}, although the
formation of a Keplerian disc required a larger Hall parameter than that
expected from the microphysics, possibly because the 10 au sink particle
prevented the outwards growth of a disc from the geometric centre of the
collapse. 

Further work must be done to study the role of the Hall effect on the magnetic
field diffusion in star formation, particularly using the semianalytic model
constructed in this work. This could include adopting more realistic values
and scalings of the ambipolar and Hall diffusion parameters, in order to
resolve the magnetic braking catastrophe; including a disc wind in place of
the cap on $B_{\phi,s}$; and exploring those regions of parameter space in
which Hall diffusion is the dominant form of field transport. 

As explained in \S\ref{ssim}, the Hall and ambipolar diffusion coefficients
in our calculations scale with the nondimensional variables in the same
manner. However, the diffusion coefficients could scale as any function of $x$
and the self-similar variables in order to mimic the behaviour expected from
ionisation equilibrium calculations. For example, the Hall parameter could
scale as 
\begin{equation}
    \frac{\eta_H'}{b} = \tilde{\eta}_H\sqrt{\frac{h}{\sigma}}\,,
\label{etahscale2}
\end{equation}
which is expected if the only particles in the collapsing core are neutrals
and ions, without grains. Such a formulation would cause Hall diffusion to
become important earlier in the collapse, and the Keplerian disc would need to
be described by a new set of asymptotic inner boundary conditions. 

The inner Keplerian disc in our solutions satisfies the criterion $B_{r,s} /
B_z > 1/\sqrt{3}$, which is the launching condition for a cold,
centrifugally-driven wind, and the radial scaling of the magnetic field
components is identical to that of the radially-self-similar wind solution of
\citet{bp1982}. Such a disc wind (described in appendix C of KK02) would be
the dominant mechanism for the vertical transfer of angular momentum from the
disc to the envelope, and must be included in future self-similar collapse
simulations in order to explore the influence a wind may have on the angular
momentum transport and the magnetic braking catastrophe, and to improve the
accuracy of the semianalytic models, as disc winds and jets occur in numerical
simulations of collapsing cores \citep[e.g.][]{t2002, ml2009, ch2010}, some of
which display the magnetic braking catastrophe. 

Further explorations of parameter space are required to more fully understand
the influence of the Hall effect on star and Keplerian disc formation. In
particular we have yet to fully separate the influence of Hall diffusion from
that of increased ambipolar diffusion in the core, to find similarity
solutions in which Hall diffusion is the dominant form of flux transport, or
solutions in which there is no ambipolar diffusion at all. As in
\citet{kls2010-2}, solutions with only Hall diffusion see the field diffusion
depend upon the $\mathbf{J} \times \mathbf{B}$ terms, so the radial diffusion
is controlled by the azimuthal field component (which may be capped), and the
azimuthal diffusion depends upon the radial component. Additional work should
also be done to confirm that larger discs form in Hall similarity solutions
where the core is initially rapidly-rotating. 

\section{Conclusions}\label{conclusions} 

This paper described a semianalytic self-similar model of the gravitational
collapse of rotating magnetic molecular cloud cores with both Hall and
ambipolar diffusion, presenting similarity solutions that showed that the Hall
effect has a profound influence on the dynamics of collapse. The solutions
satisfied the vertically-averaged self-similar equations for MHD collapse
under the assumptions of axisymmetry and isothermality, matching onto
self-similar power law relations describing an isothermal core at the moment 
of point mass formation on the outer boundary and a Keplerian disc on the
inner boundary. 

The inner solution describes a Keplerian disc in which accretion through the
disc depends upon the magnetic diffusion; with an appropriate value of the
nondimensional Hall diffusion parameter $\tilde{\eta}_H$ a stable
rotationally-supported disc forms in which the surface density $\Sigma$ scales
as $r^{-3/2}$ and vertical field strength $B_z \propto r^{-5/4}$. These are
the scalings expected from other simulations of protostellar discs to which
the solutions calculated in this work compare favourably. No disc may form
in solutions without counter-rotation when the Hall parameter is large (in
comparison to the ambipolar diffusion parameter) and has the wrong sign (which
indicates the orientation of the magnetic field with respect to the axis of
rotation), as the diffusion in these solutions is too strong and causes
disruptive torques that form subshocks in the similarity solutions. This
behaviour occurs because the response of the fluid to Hall diffusion is not
invariant under a global reversal of the magnetic field. 

The size of the rotationally-supported disc in the full similarity solutions
was shown to vary with the amount of Hall and ambipolar diffusion affecting
the pseudodisc through their effect on the magnetic braking in the fluid. By
creating an additional torque on the disc, Hall diffusion can either increase
or decrease the angular momentum and rotational support in the infalling
fluid, leading to an order of magnitude change in the Keplerian disc radius
between the similarity solutions at the extremes of $-0.5 \le \tilde{\eta}_H /
\tilde{\eta}_A \le 0.2$ (where the ambipolar diffusion parameter,
$\tilde{\eta}_A = 1$). A small amount of Hall diffusion was shown have a large
effect on the solution because the dynamic range of collapse is itself many
orders of magnitude in space and time. Hall diffusion causes there to be a
preferred handedness to the field alignment and the direction of rotation in
forming a large Keplerian disc that could be observed using next-generation
instruments such as ALMA. 

The accretion rate onto the central point mass is similarly influenced by Hall
diffusion. This is a smaller effect than that on the disc radius, as between
$\tilde{\eta}_H = \pm 0.1 \tilde{\eta}_A$ (again with $\tilde{\eta}_A = 1$)
the accretion rate onto the protostar only changes by 6 per cent, or $0.2
\times 10^{-6}$ M$_\odot$ yr$^{-1}$. There exists a clear trend in which the
protostellar accretion rate drops off with increasingly negative Hall
parameter despite the constant accretion onto the core, as the reduced
magnetic braking in these solutions causes a larger Keplerian disc to form,
and accretion through this disc onto the protostar is slow. 

The magnetic braking catastrophe could be resolved by the inclusion of Hall
diffusion in numerical solutions, as with one sign of $\tilde{\eta}_H$ the Hall
effect acts to reduce the total amount of braking affecting the core,
preventing it from removing too much angular momentum from the collapse.
However, with the other sign of $\tilde{\eta}_H$ the magnetic braking is
increased so that more angular momentum is transported to the envelope. As
magnetic braking due to Hall diffusion does not stop acting once no angular
momentum remains (as ambipolar diffusion does) it could also then spin the
collapsing fluid back up in the opposite direction to the initial rotation.
This acceleration is only possible with Hall diffusion, and it has the
potential to completely resolve the magnetic braking catastrophe. 

Because of its tendency to move the magnetic field in unusual directions Hall
diffusion is usually overlooked in simulations of gravitational collapse and
star formation. It has been shown that the Hall effect is important to the
dynamics of the collapse, particularly the magnetic braking behaviour which
determines the existence and size of the rotationally-supported protostellar
disc. The handedness of the response of the collapse to the inclusion of the
Hall effect has obvious dynamical and potentially observable consequences for
the gravitational collapse of molecular cloud cores, which must be studied
more closely if the dynamics of the star formation process and the variations
observed across YSOs and their discs are to be properly understood. 

\section*{Acknowledgments}

The authors wish to thank Ruben Krasnopolsky for his helpful comments on the
numerical problems and counter-rotation, and the participants of the Dynamics
of Discs and Planets Programme at the Isaac Newton Institute for Mathematical
Sciences for the engaging and stimulating discussions. This work was supported
in part by the Australian Research Council grant DP0881066.

\onecolumn
\appendix

\section{Vertical Averaging}\label{vertavg}

In order to produce a set of disc equations that depend only upon $r$ and $t$,
we average Equations \ref{m2}--\ref{in2} vertically over the disc in order to
reduce the dimensionality of the problem. As the equations for the
conservation of mass, radial and angular momentum and the vertical hydrostatic
balance remain unchanged from KK02, the reader is directed to their appendix A
or \S 2.3 of \citet{b2011} for the details of this averaging, and present here
only the derivations of the vertically-averaged induction equation and the
azimuthal field component. 

\subsection{$z$-component of the induction equation}

The $z$-component of the induction equation, given in Equation \ref{in2}, is
expanded more completely as
\begin{align}
    \frac{\partial{B_{z}}}{\partial{t}} 
    = - \frac{1}{r}\frac{\partial}{\partial{r}} 
    \Biggl[ r\Biggl(V_{r}B_{z} 
    &+ \eta \left(\frac{\partial{B_{r}}}{\partial{z}} 
      - \frac{\partial{B_{z}}}{\partial{r}}\right) 
    + \frac{\eta_{H}}{B}\left(B_{z}\frac{\partial{B_{\phi}}}{\partial{z}} 
      + \frac{B_r}{r}\frac{\partial}{\partial{r}}(rB_{\phi})\right)\nonumber\\
    &- \frac{\eta_{A}}{B^{2}}\biggl((B_{z}^{2} + B_{r}^{2})
      \left(\frac{\partial{B_{z}}}{\partial{r}} 
      - \frac{\partial{B_{r}}}{\partial{z}}\right) 
      - B_{r}B_{\phi}\frac{\partial{B_{\phi}}}{\partial{z}}
      + \frac{B_{z}B_{\phi}}{r}\frac{\partial}{\partial{r}}(rB_{\phi})\biggr) 
    \Biggr)\Biggr].
    \label{in3e}
\end{align}
The magnetic flux enclosed within a radius $r$ is given by
\begin{equation}
    \Psi(r) = \Psi_c + 2\pi\int^{r}_{0}B_z(r')r'dr',
\label{intpsi}
\end{equation}
where $\Psi_c$ is the flux within the central point mass. This equation is
then rewritten in differential form as
\begin{equation}
    B_z = \frac{1}{2\pi{r}}\frac{\partial{\Psi}}{\partial{r}},
\label{difpsi}
\end{equation}
and its derivative with respect to time is
\begin{equation}
    \frac{\partial{B_{z}}}{\partial{t}} 
    = \frac{1}{2\pi{}r}\frac{\partial}{\partial{r}}
    \left(\frac{\partial{\Psi}}{\partial{t}}\right).
    \label{psi}
\end{equation}
This is substituted into Equation \ref{in3e} and the partial derivative with
respect to $r$ and the factor of $r^{-1}$ are cancelled to obtain 
\begin{align}
    \frac{1}{2\pi{}}\frac{\partial{\Psi}}{\partial{t}}
    = - r\biggl[V_{r}B_{z} &+ \eta 
		\left( \frac{\partial{B_{r}}}{\partial{z}} 
            - \frac{\partial{B_{z}}}{\partial{r}}\right)
    + \frac{\eta_{H}}{B}\left(B_{z}\frac{\partial{B_{\phi}}}{\partial{z}} 
      + \frac{B_r}{r}\frac{\partial}{\partial{r}}(rB_{\phi})\right)
	\nonumber\\ 
     &- \frac{\eta_{A}}{B^{2}}\left((B_{z}^{2} + B_{r}^{2})
      \left(\frac{\partial{B_{z}}}{\partial{r}} 
      - \frac{\partial{B_{r}}}{\partial{z}}\right) 
    - B_{r}B_{\phi}\frac{\partial{B_{\phi}}}{\partial{z}}
    + \frac{B_{z}B_{\phi}}{r}\frac{\partial}{\partial{r}}(rB_{\phi})\right) 
    \biggr]. 
\label{in4}
\end{align}
The $\eta$ and $\eta_{H, A}$ terms depend on $B^{0,1,2}$ respectively, so the
leading fractions of the diffusive terms may be ignored as the integration
over $z$ is performed.  

The flux, magnetic force and Ohmic diffusion terms are integrated over the
disc height to give: 
\begin{equation}
    \int_{-\infty}^{+\infty}{\frac{1}{2\pi}}\frac{\partial\Psi}{\partial{t}}dz
    =\left[\frac{1}{2\pi}\frac{\partial\Psi}{\partial{t}}z\right]^{+H}_{-H}
     =\frac{2H}{2\pi}\frac{\partial\Psi}{\partial{t}};
\label{in5-1}
\end{equation}
\begin{equation}
    \int_{-\infty}^{+\infty}{V_rB_z}dz
    = \Biggl[V_rB_zz\Biggr]^{+H}_{-H} = 2HV_rB_z;
\label{in5-2}
\end{equation}
and
\begin{equation}
    \int_{-\infty}^{+\infty}\left(\frac{\partial{B_r}}{\partial{z}}
        - \frac{\partial{B_z}}{\partial{r}}\right)dz
    = \int_{-H}^{+H}
        \frac{\partial}{\partial{z}}\left(\frac{B_{r,s}z}{H}\right)
        - \frac{\partial{B_z}}{\partial{r}}dz
    = 2\left(B_{r,s} - H\frac{\partial{B_z}}{\partial{r}}\right).
\label{in5-3}
\end{equation} 
The Hall diffusion terms are rearranged into the form 
\begin{equation}
    B_z\frac{\partial{B_\phi}}{\partial{z}}
        + \frac{B_r}{r}\frac{\partial}{\partial{r}}(rB_\phi{})
    = \frac{\partial}{\partial{z}}(B_zB_\phi) 
            - B_\phi\frac{\partial{B_z}}{\partial{z}}
            + \frac{B_r}{r}\frac{\partial}{\partial{r}}(rB_\phi);
\label{in5-4-1}
\end{equation}
and the vertical scaling of the azimuthal field component is substituted into
the first term and the solenoidal condition (Equation \ref{solenoid}) is
applied to the second:
\begin{equation}
    B_z\frac{\partial{B_\phi}}{\partial{z}}
        + \frac{B_r}{r}\frac{\partial}{\partial{r}}(rB_\phi{})
    = \frac{\partial}{\partial{z}}\left(\frac{B_zB_{\phi,s}z}{H}\right)
            + \frac{B_\phi}{r}\frac{\partial}{\partial{r}}(rB_r) 
            + \frac{B_r}{r}\frac{\partial}{\partial{r}}(rB_\phi).
\label{in5-4-2}
\end{equation}
The integral of the Hall terms may then by written as
\begin{equation} 
    \int_{-\infty}^{+\infty}\left[B_z\frac{\partial{B_\phi}}{\partial{z}}
        + \frac{B_r}{r}\frac{\partial}{\partial{r}}(rB_\phi{})\right]dz
    = \int_{-\infty}^{+\infty}
        \left[\frac{\partial}{\partial{z}}\left(\frac{B_zB_{\phi,s}z}{H}\right)
          + \frac{1}{r^2}\frac{\partial}{\partial{r}}(r^2B_rB_\phi)\right]dz,
\label{in5-4-3}
\end{equation}
which, after the vertical scalings of $B_r$ and $B_\phi$ are substituted into
it, is evaluated to give: 
\begin{align}
    \int_{-\infty}^{+\infty}\left[B_z\frac{\partial{B_\phi}}{\partial{z}}
      + \frac{B_r}{r}\frac{\partial}{\partial{r}}(rB_\phi{})\right]dz
    &= \left[\frac{B_zB_{\phi,s}z}{H} 
      + \frac{z^3}{3r^2}\frac{\partial}{\partial{r}}
      \left(\frac{r^2B_{r,s}B_{\phi,s}}{H^2}\right)\right]^{+H}_{-H}\nonumber\\
    &= 2B_zB_{\phi,s} + \frac{2H^3}{3r^2}\frac{\partial}{\partial{r}}
      \left(\frac{r^2B_{r,s}B_{\phi,s}}{H^2}\right).
\label{in5-4-4}
\end{align}

Finally, the ambipolar diffusion terms are expanded out and integrated. The
first of these terms is straightforward, as $B_z$ is regarded as constant with
height unless specifically differentiated with respect to $z$ and may be taken
outside of the integral, which is solved to obtain
\begin{equation}
    \int_{-\infty}^{+\infty} (B_r^2+B_z^2) \frac{\partial{B_z}}{\partial{r}}dz
    = \frac{\partial{B_z}}{\partial{r}}\int_{-H}^{+H}
        \left(\frac{B_{r,s}^2z^2}{H^2} + B_z^2\right)dz 
    = 2H\frac{\partial{B_z}}{\partial{r}}
        \left(\frac{B_{r,s}^2}{3} + B_z^2\right).
\label{in5-5}
\end{equation}
The second of the ambipolar diffusion terms is rearranged into the form
\begin{equation} 
    (B_r^2+B_z^2) \frac{\partial{B_r}}{\partial{z}}
    = B_r^2\frac{\partial{B_r}}{\partial{z}} 
      + B_z\frac{\partial}{\partial{z}}(B_rB_z) 
      - B_rB_z\frac{\partial{B_z}}{\partial{z}}
\label{in5-6-1}
\end{equation}
to which the solenoidal condition (Equation \ref{solenoid}) and the scalings
for the other field components are applied. The integral of this term is then
\begin{equation}
   \int_{-\infty}^{+\infty}(B_r^2+B_z^2) \frac{\partial{B_r}}{\partial{z}} dz
   = \int_{-\infty}^{+\infty} \frac{B_{r,s}^2z^2}{H^2}
      \frac{\partial}{\partial{z}}\left(\frac{B_{r,s}z}{H}\right)
   + B_z\frac{\partial}{\partial{z}} \left(\frac{B_{r,s}B_{z}z}{H}\right)
   + \frac{B_{r,s}B_zz^2}{Hr}\frac{\partial}{\partial{r}}
     \left(\frac{rB_{r,s}}{H}\right) dz;
\label{in5-6-2}
\end{equation}
this is evaluated over the height of the disc to give
\begin{align}
    \int_{-\infty}^{+\infty}(B_r^2+B_z^2)
        \frac{\partial{B_r}}{\partial{z}}dz
    &= \left[\frac{B_{r,s}^3z^3}{3H^3} + \frac{B_z^2B_{r,s}z}{H} 
        + \frac{B_{r,s}B_zz^3}{3rH}\frac{\partial}{\partial{r}}
        \left(\frac{rB_{r,s}}{H}\right)\right]^{+H}_{-H}\nonumber\\
    &= \frac{2}{3}B_{r,s}^3  + 2B_z^2B_{r,s}
        + \frac{2}{3}H^2B_{r,s}^2B_z
        \left[\frac{d}{dr}[\ln(rB_{r,s})] - \frac{d}{dr}[\ln{H}]\right].
    \label{in5-6-3}
\end{align}
The third of the ambipolar diffusion terms is again straightforward; it is
vertically-averaged by applying the vertical scalings to the radial and
azimuthal components to the field and then performing the integral over $z$ to
find 
\begin{equation}
    \int_{-\infty}^{+\infty}B_rB_\phi\frac{\partial{B_\phi}}{\partial{z}}dz
    = \int_{-\infty}^{+\infty}\frac{B_{r,s}B_{\phi,s}z^2}{H^2}
      \frac{\partial}{\partial{z}}\left(\frac{B_{\phi,s}z}{H}\right)dz
    = \frac{2}{3}B_{r,s}B_{\phi,s}^2.
\label{in5-7}
\end{equation}
Finally, the last of the ambipolar diffusion terms in Equation \ref{in4} is
averaged by substituting in the vertical scalings of the field components
and then performing the integral:
\begin{align}
    \int_{-\infty}^{+\infty}\frac{B_\phi{}B_z}{r}
        \frac{\partial}{\partial{r}}(rB_\phi)dz
        &= \int_{-\infty}^{+\infty}\frac{B_{\phi,s}B_zz}{rH}
        \frac{\partial}{\partial{r}}\left(\frac{rB_{\phi,s}z}{H}\right)dz\nonumber\\
    &= \frac{2}{3}B_zB_{\phi,s}^2H
        \left[\frac{d}{dr}[\ln(rB_{\phi,s})] - \frac{d}{dr}[\ln{H}]\right].
\label{in5-8}
\end{align}

Collecting all of these integrated terms into the same order as in Equation
\ref{in4} then gives the full vertically-averaged induction equation: 
\begin{align}
    \frac{H}{2\pi}\frac{\partial\Psi}{\partial{t}}
    = -r\Biggl[HV_rB_z 
     &+ \eta\!\left(\!B_{r,s} - H\frac{\partial{B_z}}{\partial{r}}\!\right)\!
     + \frac{\eta_H}{B} \!\left(\!B_zB_{\phi,s} + \frac{H^3}{3r^2}
       \frac{\partial}{\partial{r}}
       \!\left(\!\frac{r^2B_{r,s}B_{\phi,s}}{H^2}\!\right)\!\right)\nonumber\\
    &- \frac{\eta_A}{B^2}\!\biggl[\!
       -\!\left(\!B_{r,s} - H\frac{\partial{B_z}}{\partial{r}}\!\right)\!
       \!\left(B_z^2 + \frac{1}{3}B_{r,s}^2\right)\!
       - \frac{1}{3}B_{\phi,s}^2B_{r,s}    \label{in6}\\
    &+ \frac{1}{3}HB_zB_{\phi,s}^2
       \!\left(\!\frac{d}{dr}[\ln(rB_{\phi,s})] 
       - \frac{d}{dr}[\ln{H}]\!\right)\! - \frac{1}{3}HB_zB_{r,s}^2
	\left(\frac{d}{dr}[\ln(rB_{r,s})] - \frac{d}{dr}[\ln{H}]\!\right)\!
        \!\biggr]\!\Biggr] .\nonumber
\end{align}
It is clear from this equation that the azimuthal field is pivotal in causing
Hall drift in the radial direction; $B_{\phi,s}$ should not be neglected, even
in axisymmetric models.

\subsection{Azimuthal field component}

The vertical angular momentum transport above and within the pseudodisc is
achieved by magnetic braking, especially during the dynamic collapse phase
inwards of the magnetic diffusion shock. It is assumed that magnetic braking
remains the dominant angular momentum transport mechanism during the
subsequent evolution of the core, although it is likely that a
centrifugally-driven disc wind may dominate in the innermost Keplerian disc.
The approach to modelling the magnetic braking adopted here is adapted from
that of \citet{bm1994} for the pre-point mass formation collapse phase. This
formulation is not well-defined in the innermost rotationally-supported
regions of the disc, where the calculated magnetic braking becomes stronger
than is expected and the angular momentum transport is expected to be
dominated by a disc wind (this is discussed in more detail in
\S\ref{discuss}). A cap is then placed upon the azimuthal magnetic field 
component in order to ensure that it does not greatly exceed the vertical
component; because of this the magnetic braking prescription is not expected
to introduce significant errors into the inner regions of the calculations. 

External to the pseudodisc the magnetic field is considered to be frozen into
the low-density, constant-pressure external medium, which has density
$\rho_\text{ext}$ and angular velocity $\Omega_b$. Within the external medium the
magnetic field assumes the value $\mathbf{B} = B_\text{ref} \hat{z}$, and the
exterior flux tubes corotate with the core. Because the transition region has
a low moment of inertia relative to the core, and the crossing time for
Alfv\'en waves is always much smaller than the evolutionary time of the core,
the transition region can relax to a steady state during all stages of
contraction \citep{bm1994}. 

The induction equation under IMHD implies  
\begin{equation}
    (B_p \cdot \nabla) \Omega = 0
\label{Bp}
\end{equation}
where $B_p$ is the poloidal field, so that the angular velocity $\Omega$ is
constant on a magnetic surface. The force equation is similarly  
\begin{equation}
    (B_p \cdot \nabla) rB_\phi = 0,
\label{Bp2}
\end{equation}
which further implies that $rB_\phi$ does not change along the field lines.
The neutral particles carry the torque and angular momentum is carried upwards
by torsional Alfv\'en waves generated by the rotation of the disc. 

Over a period of time $dt$ an amount of material equal to $2\pi\rho\,
r_\text{ref} dr_\text{ref}$ moves from the undisturbed position $r_\text{ref}
dr_\text{ref}$ in the external medium along a flux tube with angular velocity
$\Omega$ to a radius $rdr$ at the disc surface. The angular momentum of the
gas goes as 
\begin{equation}
    dJ = - [2\pi\rho_\text{ext}{r_\text{ref}}dr_\text{ref}] (V_\text{A,ext}dt)
	r_\text{ref}^2 (\Omega - \Omega_b),
\label{dJ1}
\end{equation}
where $V_\text{A,ext}$, the external Alfv\'{e}n speed, is given by
\begin{equation}
    V_\text{A,ext} = \frac{B_\text{ref}}{\sqrt{4\pi{\rho_\text{ext}}}}.
\label{valf}
\end{equation}
For purely azimuthal motions in the external medium, the total angular
momentum in each flux tube is conserved. This angular momentum must be
removed from the disc at a rate equal to  
\begin{equation}
    \frac{dJ}{dt} = -2\pi{r_\text{ref}^2}V_{A,ext}\rho_\text{ext}(\Omega - 
       \Omega_b)r_\text{ref}dr_\text{ref},
\label{dJdt}
\end{equation}
which gives a torque on the disc
\begin{equation}
    N = -\frac{2\pi{r_\text{ref}^2}(V_\text{A,ext}\rho_\text{ext})(\Omega -
	\Omega_b) r_\text{ref}dr_\text{ref}}{\pi{rdr}}.
\label{n1}
\end{equation}
The amount of flux remains constant along flux tubes, so that the flux through
the disc inside of a radius $r$ is equal to the amount of flux through the
cylindrical external cloud inside of the radius $r_\text{ref}$:
\begin{equation}
    \Psi = \int^{r}_{0}2\pi{r'}B_{z,eq}(r')dr' 
         = \pi{r_\text{ref}^2}B_\text{ref},
\label{psi2}
\end{equation}
where $B_{z,eq}$ is the value of $B_z$ at the midplane of the disc.
Thus 
\begin{equation}
    d\Psi = 2\pi{r}B_{z,eq}dr = 2\pi{r_\text{ref}}B_\text{ref}dr_\text{ref}
\label{dpsi}
\end{equation}
and
\begin{equation}
    \frac{r_\text{ref}dr_\text{ref}}{rdr} = \frac{B_{z,eq}}{B_\text{ref}},
\label{bzeq}
\end{equation}
so that the torque in Equation \ref{n1} becomes
\begin{equation}
    N =-\frac{2r^2_\text{ref}(\Omega - \Omega_b)B_{z,eq}}{B_\text{ref}}
         \left(\frac{B_\text{ref}\rho_\text{ext}}{\sqrt{4\pi\rho_\text{ext}}}\right)
      =-\frac{(\Omega - \Omega_b)B_{z,eq}(\Psi/2\pi)}{\pi{}V_\text{A,ext}}.
\label{n1-2}
\end{equation}

The torque per unit area on the disc is given by
\begin{equation}
    N = \frac{rB_{z,eq}B_{\phi,s}}{2\pi};
\label{n2}
\end{equation}
combining Equations \ref{n1-2} and \ref{n2} gives the steady state azimuthal
magnetic field component at the surface of the disc: 
\begin{equation}
    B_{\phi,s} = -\frac{\Psi}{\pi{r^2}}\frac{(r\Omega -
r\Omega_b)}{V_\text{A,ext}}
\label{b_phisdef1}
\end{equation}
(equation 26 of Basu \& Mouschovias, 1994; equation 3 of KK02). It is clear
that the properties of the external medium determine the conditions at the
disc surface. This steady state approximation requires that the ratio of the
Alfv\'{e}n travel time in the external medium to the initial radius of the
cloud be less than the evolutionary timescale, which scales with $r$ as $\sim
r/|V_r|$. For the rotationally-supported discs presented here $|V_r| \lesssim
c_s$ (and $|V_r| \to 0$ as $r \to 0$), which is much smaller than the Alfv\'en
speed for the adopted gas temperature of 10K ($V_\text{A,ext} \approx 5c_s$);
this implies that the assumption of rapid braking of the core should not
introduce large errors into the solutions. 

The angular velocity $\Omega$ is given by the equation
\begin{equation}
    \Omega = \frac{1}{r}(V_\phi + V_{B\phi}),
\label{omega}
\end{equation}
where, using $\eta_P = \eta_A + \eta$,
\begin{equation} 
    V_{B\phi} = -\frac{1}{B}\left[\eta_H(\nabla\times\mathbf{B})_\perp 
      -\eta_P(\nabla\times\mathbf{B})_\perp\times\mathbf{\hat{B}}\right]_\phi.
\label{vbphi1}
\end{equation}
This equation is then expanded out to become
\begin{align}
    V_{B\phi} &= - \frac{1}{B^2}\Biggl[\frac{\eta_H}{B}
      \left((B_z^2+B_r^2)\left(\frac{\partial{B_z}}{\partial{r}}
        -\frac{\partial{B_z}}{\partial{z}}\right) 
        + B_{\phi}B_z\frac{1}{r}\frac{\partial}{\partial{r}}(rB_\phi)
        - B_{\phi}B_r\frac{\partial{B_\phi}}{\partial{z}}\right)\nonumber\\
    &\qquad \qquad - \frac{\eta_P}{B^2}(B_r^2 + B_\phi^2 + B_z^2)
        \left(\frac{B_r}{r}\frac{\partial}{\partial{r}}(rB_\phi)
        + B_z\frac{\partial{B_\phi}}{\partial{z}}\right) \Biggr],
\label{vbphi2}
\end{align}
Most of the terms in this equation have direct analogies in Equation
\ref{in4}, and the individual steps of the vertical integration are not
reproduced here. The vertical averaging gives 
\begin{align}
    HV_{B\phi} \biggl(\!\frac{1}{3}B_{r,s}^2 + \frac{1}{3}B_{\phi,s}^2 
      + B_z^2\!\biggr) = &- \frac{\eta_H}{B} \Biggl[\! 
    \left(\!\frac{B_{r,s}^2}{3} + B_z^2\!\right)\!\! \left(\!B_{r,s} 
     - H\frac{\partial{B_z}}{\partial{r}}\!\right) + \frac{1}{3}HB_zB_{r,s}^2
     \!\left(\!\frac{d}{dr}[\ln(rB_{r,s})] - \frac{d}{dr}[\ln{H}]\!\right)
    \nonumber\\ 
    &\quad \quad+ \frac{1}{3}B_{r,s}B_{\phi,s}^2 
     - \frac{1}{3}HB_zB_{\phi,s}^2 \!\left(\!\frac{d}{dr}[\ln(rB_{\phi,s})] 
     - \frac{d}{dr}[\ln{H}]\!\right)\!\Biggr]\! \nonumber\\
    &+\frac{\eta_P}{B^2} \Biggl[\!B_zB_{\phi,s}
       \left(\!\frac{1}{3}B_{r,s}^2 + \frac{1}{3}B_{\phi,s}^2 + B_z^2\!\right)
    \nonumber\\ &\quad \quad+ HB_{r,s}B_{\phi,s}
    \biggl(\!\frac{B_{r,s}^2}{5} + \frac{B_{\phi,s}^2}{5} 
     + \frac{B_z^2}{3}\!\biggr)\!\! \biggl(\!\frac{d}{dr}[\ln(rB_{\phi,s})] 
     + \frac{d}{dr}[\ln(rB_{r,s})] - 2\frac{d}{dr}[\ln{H}]\!\biggr) 
    \!\Biggr]\!. \label{vb2}
\end{align}
This equation is simplified as in \S\ref{equations} by omitting any terms of
order ${\cal O}(H/r)$ save for the $[B_{r,s} - H(\partial{B_z}/\partial{r})]$
term; and the final form of $V_{B\phi}$ is then: 
\begin{equation}
    V_{B\phi} = -\frac{1}{H}\left[\frac{\eta_H}{B}\left(B_{r,s} 
	- H\frac{\partial{B_z}}{\partial{r}}\right) 
	- \frac{\eta_P}{B^2}B_zB_{\phi,s}\right].
\label{vb3}
\end{equation}
This is equivalent to the ion-neutral drift velocity adopted by KK02 (their
equation 9), with the inclusion of terms describing the effect of Hall
diffusion. 

The $\Omega_b$ term is dropped from Equation \ref{b_phisdef1}, as the
molecular cloud rotation rate is slow compared with that of the collapsing
material. Rotation is dynamically important in the inner regions of the
solutions presented in this thesis, while it is not important in most
molecular clouds, so it is reasonable to declare that $\Omega \gg \Omega_b$
and dismiss $\Omega_b$ as small. The external Alfv\'en speed,
$V_\text{A,ext}$, is treated as a constant with respect to the isothermal
sound speed in these calculations, parameterised by the constant $\alpha$
(defined in Equation \ref{alpha}). This scaling of $V_\text{A,ext}$ is
reasonable as the observations by \citet{c1999} indicated that $V_\text{A}
\approx 1$ km s$^{-1}$ over at least four orders of magnitude in density
($\sim 10^3$--$10^7$ cm$^{-3}$) in their observed molecular clouds. 

Equations \ref{vb3} and \ref{alpha} are substituted into \ref{b_phis1} to
find that
\begin{equation}
    B_{\phi,s} = -\frac{\Psi\alpha}{\pi{r^2}c_s}
     \left[\frac{J}{r} - \frac{\eta_H}{B}
       \left(B_{r,s} - H\frac{\partial{B_z}}{\partial{r}}\right)\right]
     \left[1 + \frac{\Psi\alpha}{\pi{r^2}c_s}\frac{\eta_P}{B^2}
       \frac{B_z}{H}\right]^{-1}.
\label{b_phis}
\end{equation}
Note that $B$ has an implied $B_{\phi,s}$ dependence; this is typically solved
for numerically when calculating the azimuthal field. 

For the inner solutions, $\Omega$ increases with decreasing $r$ (proportional
to $r^{-3/2}$); this would make $B_{\phi,s}$ the dominant field component at the
surface near to the central point mass. Such behaviour is not expected in a
real disc, where internal kinks of the field and magnetohydrodynamical
instabilities (for example, the magnetorotational instability) should reduce
the value of $B_{\phi}$ at the surface. An artificial limit on $B_{\phi,s}$ is
imposed:  
\begin{equation}
|B_{\phi,s}| \leq \delta{B_z}, 
\label{b_phislim}
\end{equation}
where $\delta$ is a parameter of the model usually chosen to be $\delta = 1$
in order to ensure that the azimuthal field component does not exceed the
vertical component. KK02 point out that this value quite conveniently
corresponds to that expected for a rotationally-supported disc where the
vertical angular momentum transport is dominated by a centrifugally-driven
wind. Applying this cap to Equation \ref{b_phis} then gives the final equation
for $B_{\phi,s}$:
\begin{equation}
    B_{\phi,s} = -\min\left[\!\frac{\Psi\alpha}{\pi{r^2}c_s}
     \left[\frac{J}{r} - \frac{\eta_H}{B}\!
       \left(\!B_{r,s} - H\frac{\partial{B_z}}{\partial{r}}\!\right)\!\right]
     \!\!\left[1 + \frac{\Psi\alpha}{\pi{r^2}c_s}\frac{\eta_P}{B^2}
       \frac{B_z}{H}\right]^{-1};\delta{B_z}\right].
\label{b_phisfinal}
\end{equation}

\bsp
\label{lastpage}

\end{document}